\documentclass[%
reprint,
amsmath,amssymb,
prb
]{revtex4-1}
\usepackage{graphicx}
\usepackage{dcolumn}
\usepackage{bm}
\usepackage{hyperref}
\usepackage[mathlines]{lineno}
\usepackage[caption=false]{subfig}
\usepackage{amsmath}
\usepackage{amssymb}
\usepackage{array}
\def\beq{\begin{equation}}
\def\eeq{\end{equation}}
\def\bea{\begin{eqnarray}}
\def\eea{\end{eqnarray}}
\def\k{{\bf k}}
\def\p{{\bf p}}
\def\q{{\bf q}}
\def\J{\hat{\mathbf{J}}}
\def\r{{\hat \rho}}
\def\cs{^{c(s)}}
\def\nn{\nonumber}
\def \tq{q}
\def \tp{p}
\def \tk{k}

\renewcommand\bm{\mathbf}

\begin{document}

\title{The Conditions for $l=1$ Pomeranchuk Instability in a Fermi Liquid}
\author{Yi-Ming Wu}
\author{Avraham Klein}
\author{Andrey V. Chubukov}
\affiliation{School of Physics and Astronomy, University of Minnesota, MN}

\date{\today}

\begin{abstract}
  We perform a microscropic analysis of how the constraints imposed by conservation laws affect $q=0$ Pomeranchuk instabilities in a Fermi liquid. The conventional view is that these instabilities are determined by the static interaction between low-energy quasiparticles near the Fermi surface, in the limit of vanishing momentum transfer $q$. The condition for a Pomeranchuk instability is set by $F\cs_l =-1$, where  $F\cs_l$ (a Landau parameter) is a properly normalized partial component of the anti-symmetrized static interaction $F(k,k+q; p,p-q)$ in a charge (c)  or spin (s) sub-channel with angular momentum $l$.  However, it is known that conservation laws for total spin and charge prevent Pomeranchuk instabilities for $l=1$ spin- and charge- current order parameters. Our study aims to understand whether this holds only for these special forms of $l=1$ order parameters, or is a more generic result. To this end we perform a diagrammatic analysis of spin and charge susceptibilities for charge and spin density order parameters, as well as perturbative calculations to second order in the Hubbard $U$. We argue that for $l=1$ spin-current and charge-current order parameters, certain vertex functions, which are determined by high-energy fermions, vanish at $F\cs_{l=1}=-1$, preventing a Pomeranchuk instability
  from taking place.
  For an order parameter with a generic $l=1$ form-factor, the vertex function is not expressed in terms of  $F\cs_{l=1}$, and a Pomeranchuk instability does occur when $F\cs_1=-1$. We argue that for other values of $l$,
  a Pomeranchuk instability occurs at  $F\cs_{l} =-1$
  for an order parameter
  with any form-factor
\end{abstract}

\maketitle

\section{Introduction}
\label{sec:introduction}
 This  paper is devoted to the analysis of subtle effects associated with a Pomeranchuk instability in a Fermi liquid (FL) due to  the interplay with conservation laws.
 A system of interacting fermions is called a Fermi liquid  if its properties differ from those of free fermions in a quantitative, but not qualitative manner \cite{AGD,Lifshitz1980,Shankar1994}. Specifically,  the distribution function  $n_k$  undergoes a finite jump at the Fermi momentum, $k_F$,  with some jump magnitude $Z < 1$; the velocity $v^*_F$  of fermionic excitations near the Fermi surface (FS)  remains finite; and the lifetime of fermionic excitations near a FS is parametrically larger than the energy counted from the Fermi level,  i.e., fermions infinitesimally close to the FS can be viewed as infinitely long lived.  These three features form the basis for the description of low-energy fermionic states in terms of quasiparticles, whose distribution function at $T=0$ is a step function. The validity  of FL postulates has been verified in microscopic calculations for realistic interaction potentials and was found to hold at small/moderate couplings in dimensions $d >1$.

 Stronger interactions can destroy a FL. In general, such destruction can occur in two ways.
 One option is the transformation of a metal into a Mott insulator,  once the interaction $U$ becomes comparable to a fermionic bandwidth $W$. This instability  involves fermions located everywhere in the Brillouin zone.  Another option is an instability driven by fermions only very near the FS, such as superconductivity and $q=0$  instabilities in a particle-hole channel, often called Pomeranchuk instabilities. The latter leads to either phase separation, or ferromagnetism, or a deformation of a FS and the development of a particle-hole order with non-zero angular momentum
 (see e.g. Refs \onlinecite{Pethick1988,Moriya1985,Metzner2003,Oganesyan2001,Maslov2010,Halboth,Oganesyan2001,Metzner2003,Quintanilla,Senthil2008,Hirsch,Chubukov2009}).
  A Pomeranchuk instability in a given channel occurs when the corresponding interaction exceeds $1/N_F$, where $N_F$ is the  density of states at the FS. When $W N_F \gg 1$, a Pomeranchuk instability occurs well inside the metallic regime.

  A Pomeranchuk instability is generally expressed as a condition on  a Landau parameter.
For a rotationally-invariant and SU(2) spin-invariant FL, an anti-symmetrized static interaction between fermions at the FS and at strictly zero momentum transfer, $\Gamma^\omega (k,k;p,p)$ can be separated into spin and charge components, and each can be further decomposed into sub-components with different angular momenta $l$. Landau parameters are properly normalized dimensionless sub-components $F_{l}^{c(s)}$, where  $c(s)$ selects charge (spin) channel, and  $l=0,1,2,...$  ~\cite{Landau1,Landau2,AGD,Lifshitz1980}. Pomeranchuk argued in his original paper \cite{Pomeranchuk} that a static susceptibility $\chi^{c(s)}_{l}$ scales as $1/(1 + F^{c(s)}_l)$ and diverges when the corresponding $F^{c(s)}_l = -1$. The divergence  signals an instability towards a $q=0$ density-wave order with angular momentum $l$.

The $1/(1 + F^{c(s)}_l)$ form of the susceptibility can be reproduced diagrammatically by summing up particle-hole bubbles of free fermions within RPA. The momentum/frequency integration within each bubble is confined to the FS, hence the dimensionless interaction between the bubbles is exactly $F^{c(s)}_l$. The RPA series are geometric, hence $\chi^{c(s)}_{l, RPA} =  \chi_{l,0} /(1 + F^{c(s)}_l)$, where $ \chi_{l,0}$ is a free-fermion susceptibility. This agrees with the exact forms of the susceptibilities of the $l=0$ order parameters, which correspond to the total charge and the total spin:
\begin{equation}
  \chi^{c(s)}_{l=0} = \chi_{l=0,0} \frac{m^*/m}{1 + F^{c(s)}_{l=0}},
  \label{n1}
\end{equation}
where $m^* = k_F/v^*_F$.
It is tempting to assume
 that RPA works for a generic order parameter with angular momentum $l$.  However, corrections to RPA are of order one when $F^{c(s)}_l = O(1)$, and it is a'priori unclear whether in the generic case the full susceptibility has the same functional form as $\chi^{c(s)}_{l, RPA}$.

One can actually go beyond RPA and obtain the exact expression for a static $\chi^{c(s)}_l$ for a generic order parameter
   \begin{align}\label{eq:rho-def}
    \r^{c}_l (\q) &= \sum_{\k,\alpha} \lambda^c_l (k) c^{\dagger}_{\k-\q/2,\alpha}c_{\k+\q/2,\alpha}, \\
    {\bf \r}^{s}_l (\q) &= \sum_{\k,\alpha\beta}  \lambda^s_l (k)  c^{\dagger}_{\k-\q/2,\alpha} {\bf \sigma}^{\alpha\beta}c_{\k+\q/2,\beta},
   \end{align}
   with any $l$ and any form-factor  $\lambda^{c(s)}_l (k)$. The exact formula
   was originally obtained
   by Leggett \cite{Leggett1965}, based on earlier work by Eliashberg~\cite{Eliashberg1962} (we present the diagrammatic derivation in Sec. \ref{sec:2}).
   It reads
   \begin{equation}
	 	\chi_{l}^{c(s)}= \left(\Lambda^{c(s)}_l Z\right)^2  \chi^{c(s)}_{l,qp}+\chi^{c(s)}_{l,inc}.
\label{eq:full}
 \end{equation}
 Here
 \begin{equation}
 \chi^{c(s)}_{l,qp} = \frac{m^*}{m}  \chi^{c(s)}_{l,RPA}
 \label{eq:full_1}
 \end{equation}
 is the RPA result with
 an extra factor of
 $m^*/m$, often called the quasiparticle contribution.
 Other terms in Eq. \eqref{eq:full} describe two contributions that incorporate effects beyond RPA.
 First,
 $\chi^{c(s)}_{l,qp}$ gets multiplied by $(\Lambda^{c(s)}_l Z)^2$,  where $Z$ is the quasiparticle residue and $\Lambda^{c(s)}_l$ accounts for the renormalization of the vertex containing the form-factor.
 Second,
 there is an extra term $\chi^{c(s)}_{l,inc}$. These additional terms come from processes in which at least one fermion is located away from the FS.

Although the $Z$-factor itself comes from fermions away from the FS, its presence in (\ref{eq:full}) can be easily understood because
the fermionic propagator near the FS is
 \begin{equation}
G (\omega, k)  \approx Z G_{qp} (\omega,k) =  \frac{Z}{\omega - v^*_F (k-k_F) + i \delta {\text sgn} \omega},
 \label{G}
 \end{equation}
 hence there is a $Z^2$ factor in each bubble ($Z^2$ also appears in the normalization of $F^{c(s)}_l$, see below). As long as $Z$ is a number $0<Z<1$, it alone does not change the functional form of $\chi^{c(s)}_{l}$ compared to the RPA result. The other two terms are potentially more relevant. First, the vertex function $\Lambda^{c(s)}_l$ may cancel the $1 + F^{c(s)}_l$ term and, second, either $\Lambda^{c(s)}_l$ or $\chi^{c(s)}_{l,inc}$ may diverge on their own  and give rise to a Pomeranchuk-type instability not associated with $F^{c(s)}_l = -1$ (and with the fermions on the FS).

 Some of these issues were addressed by Landau and Pitaevskii (see e.g. Ref. \onlinecite{AGD}) and by Leggett (Ref.~\onlinecite{Leggett1965}) in the early years of FL theory, by invoking (i) conservation laws and the corresponding Ward identities~\cite{AGD,Lifshitz1980} and  (ii) the continuity equation and the longitudinal sum rule~\cite{Leggett1965,Ehrenreich1967}. For a conserved order parameter, conservation laws  require that the full susceptibility coincides with the coherent term, i.e., the corresponding $\Lambda Z = 1$ and $\chi_{inc} =0$. The $l=0$ charge and spin Pomeranchuk order parameters $\r^c_{l=0}$  and ${\bf \r}^s_{l=0}$ with a {\it constant} form-factor are conserved quantities, hence the corresponding $\chi^{c(s)}_{l=0} =  \chi^{c(s)}_{l=0,qp}$, as in Eq. (\ref{n1}).  This is fully consistent with RPA. For $l=1$ charge- or spin-current order parameters with $\lambda^{c(s)}_{l=1} (k) = {\bf k}$ the continuity equation imposes the relation
 \begin{equation}
   Z\Lambda^{c(s)}_{l=1} = \frac{m}{m^*} \left(1 + F^{c(s)}_{l=1}\right)
   \label{n2}
 \end{equation}
 such that $(Z\Lambda^{c(s)}_{l=1})^2 \chi^{c(s)}_{l,qp} \propto  \left(1 + F^{c(s)}_{l=1}\right)$ vanishes at $F^{c(s)}_{l=1}=-1$ instead of diverging. In addition, the longitudinal sum rule  yields
 \begin{equation}\label{n22}
 	\chi^{c(s)}_{l=1} = \chi_{1,0},
 \end{equation}
 i.e. all interaction-induced renormalizations of $\chi^{c(s)}_{l=1}$ cancel out,
 implying,
 \begin{equation}
 \chi^{c(s)}_{l=1,inc} = \chi_{l=1,0} \left(1 - \frac{m}{m^*} \left(1 + F^{c(s)}_{l=1}\right)\right)
\label{n3}
 \end{equation}
 We emphasize that this holds even in the presence of a lattice potential $V(r)$.  For a Galilean-invariant FL,  $m^*/m$ by itself is expressed via Landau parameters as $m^*/m = 1 + F^c_{1}$.  Then, for spin-current susceptibility, $Z\Lambda^{s}_{l=1} =(1+F^s_1)/(1+ F^c_1)$ and
 $\chi^{s}_{l=1,inc}= \chi_{l=1,0} (F^c_1 -F^s_1)/(1+ F^c_1)$,
 while
 for charge-current susceptibility, $Z\Lambda^{c}_{l=1} =1$ and $\chi^{c}_{l=1,inc}=0$.  This last result is consistent with the fact that for a Galilean-invariant FL, charge-current coincides with the momentum and is a conserved quantity.

Eqs. \eqref{n2}+\eqref{n3} represent a qualitative breakdown of RPA for spin and charge-current order parameters. Within RPA, $\Lambda Z = 1$, and so to reproduce Eqs. \eqref{n2}+\eqref{n3} one must require $m^*/m = (1+ F_1^c)^{-1} = (1+F_1^s)^{-1}$. Such behavior comes about naturally if one assumes~\cite{Quintanilla} that the dressed interaction remains a function of ${\bf k} - {\bf p}$, i.e., $\Gamma^\omega(k, k;p, p) = U_{eff} (|{\bf k} - {\bf p}|)$. In this situation the charge component of $\Gamma^\omega(k, k;p, p)$ is $(U(0) - U(k-p)/2$ and the spin component is $- U(k-p)/2$.
Then  $F_l^c =  F_l^s$, for all $l > 0$, including $l=1$.
However, in fact, at order $U^2$ and higher,  the interaction gets renormalized in both particle-hole and particle-particle channels, and the renormalized interaction between fermions on the FS depends on both ${\bf k} - {\bf p}$ and ${\bf k} + {\bf p}$.  The terms which depend on ${\bf k} - {\bf p}$ and on ${\bf k} + {\bf p}$ behave differently under antisymmetrization, and, as the consequence, spin and charge components of  $\Gamma^\omega(k, k;p, p)$ are generally not equivalent for any $l$.  In this situation, Eqs. \eqref{n2} and \eqref{n3} are obeyed not because of some some special relation between Landau parameters, but rather because $\Lambda Z$ is expressed via the particular combination of Landau parameters, such that for $l=1$ spin current, $\Lambda^s_{l=1} Z$ cancels out $1+F^s_1$.

This issue has been recently re-analyzed by Kiselev et al\cite{Kiselev2017}. They discussed  how the absence of $l=1$ Pomeranchuk instability  for the spin-current order parameter  places additional  constraints on spontaneous generation of spin-orbit coupling\cite{Wu1,Wu2}, often associated with $l=1$ spin Pomeranchuk order. Kiselev et al also derived a general formula for the susceptibility of a current of a conserved order parameter.

The purpose of the current work is three-fold. First, we provide a transparent diagrammatic derivation of Eq. (\ref{eq:full}) and extend it to the case when both $q$ and $\Omega$ are small, but the ratio $v_F q/\Omega$ is arbitrary. Second, we analyze Eqs. (\ref{n2}) and (\ref{n3}) from a microscopic perspective, and identify what relates the contributions to the susceptibility from fermions near the FS,  which determine Landau parameters, and fermions  away from the FS, which  determine $Z$, $\Lambda^{c(s)}_{l}$, and $\chi^{c(s)}_{l,inc}$ (and $m^*/m$ in the absence of Galilean invariance).
Lastly, we investigate how generic is the statement about the absence of  Pomeranchuk instabilities for $l=1$ order parameters, and what happens for other $l$.

To derive Eqs. (\ref {eq:full}) and (\ref{eq:full_1})  diagrammatically, we use the expansion in the number of fermionic loops, and at each loop order separate the contributions from fermions at the FS and away from it. The contributions away from the FS can be computed by setting $q$ to zero, while for the contributions from the vicinity of the FS one needs to keep ${\bf q}$ small but finite, because each bubble contribution to $\chi^{c(s)}_{l,qp}$ comes from the tiny range near the FS where the poles of the two Green's function in a bubble are in different half-planes of frequency.  We then re-arrange the perturbation series and evaluate partial contributions to the susceptibility with $M =0,1,2$ etc. cross-sections in which the contribution comes from the FS. Summing up terms with all $M$ we reproduce Eqs. (\ref{eq:full}) and (\ref{eq:full_1}). We then extend the analysis and consider the dynamical susceptibility $\chi^{c(s)}_{l} ({\bf q}, \Omega)$ in the limit when both $|{\bf q}|$ and $\Omega$ are small, but the ratio $v^*_F |{\bf q}|/\Omega$ is arbitrary.  We show that in
an
arbitrary FL, the form of the dynamical susceptibility is rather complex, except for special cases when  $v^*_F |{\bf q}|/\Omega$ is either small or large, or  $v^*_F |{\bf q}|/\Omega$ is arbitrary, but only a few Landau parameters are not small.

In order to understand Eqs. (\ref{n2}) and (\ref{n3}) for current order parameters, we  explicitly compute low-energy and high-energy components of charge and spin susceptibilities for $l=1$ for the 2D Hubbard model, to second order in Hubbard $U$. At this order the dressed interaction between fermions becomes dynamical, and both low-energy and high-energy contributions are non-zero. For simplicity, in this calculation we neglect the lattice potential, i.e., consider a Galilean-invariant system. We show that there exists a particular identity on the sum of dynamical polarization bubbles in particle-hole and particle-particle channels, which relates the contribution to this sum coming from fermions at the FS and the one from fermions away from the FS. We use this identity to prove diagrammatically Eqs. (\ref{n2}) and (\ref{n3}). We do the same computation for $l=0$  and verify that for conserved spin and charge order parameters $\chi^{c(s)}_{l,inc} =0$ and $\Lambda^{c(s)}_l =1/Z$, i.e., $\chi^{c(s)}_{l} = \chi^{c(s)}_{l,qp}$.

We next consider $l=2$ and investigate an argument by Kiselev et al ~\cite{Kiselev2017} that for certain order parameters with $l=2$, spin and charge susceptibilities again do not diverge when the corresponding $F^{c(s)}_{l=2} =-1$ because $(1 + F^{c(s)}_{l=2})^{-1}$ in $\chi^{c(s)}_{l=2,qp}$ is canceled out by $Z \Lambda^{c(s)}_{l=2}$.  Our
perturbative
results do not support this claim. We argue that  $Z\Lambda^{c(s)}_{l=2}$ cannot be expressed solely in terms of Landau parameters. Of particular interest here is the $l=2$ charge order parameter in a Galilean-invariant case. It is tempting to view this order parameter as a current of conserved $l=1$ total momentum, and  relate the corresponding $Z \Lambda^c_{l=2}$ to Landau parameters via the continuity equation. However, we show that the current operator for momentum cannot be expressed solely in terms of bilinear combination of fermions, and contains an interaction-induced four-fermion term. As a result, the $l=2$ charge susceptibility is only a portion of the full  current-current correlator, and as such  is not determined by the continuity equation.

To understand how generic is the statement about the absence of  Pomeranchuk instabilities for $l=1$, we notice that there exists an infinite set of Pomeranchuk order parameters in any channel, including $l=0$. These order parameters contain form-factors $\lambda^{c(s)}_l (k)$, which are obtained by  multiplying the base form factor  (a constant for $l=0$, ${\bf k}$ for $l=1$,  etc), by an arbitrary function  $f_l(|{\bf k}|)$. When $f_l (|{\bf k}|)$ is not a constant, it changes the contribution to susceptibility from fermions away from the FS compared to that from fermions at the FS.   We argue that the identity, which allowed us to express $Z\Lambda^{c(s)}_{l=1}$ in terms of Landau parameters and cancel $1/(1+ F^{c(s)}_{l=1})$, does not hold if $f_l (|{\bf k}|)$ is not a constant.  As a result, $Z\Lambda^{s}_{l=1}$ no longer vanishes when $F^{s}_{l=1} =-1$. We show this non-cancellation explicitly to second order in the Hubbard $U$ by comparing susceptibilities for order parameters with form-factors ${\bf k}$ and $k_F \hat k$, where $\hat k$ is a unit vector directed along ${\bf k}$. We further argue that for a non-constant $f_l(|{\bf k}|)$, the incoherent contribution to susceptibility cannot be expressed via Landau parameters for any $l$, including $l=0$, and can potentially diverge on its own, even if the corresponding Landau parameter is still larger than $-1$.  This opens up a possibility for an instability of a FL, not associated with the singularity in the coherent part of the susceptibility.

The paper is organized as follows.   In the next Section we review the diagrammatic formulation of FL theory and present our diagrammatic derivation of Eq. (\ref{eq:full}) for the static susceptibility $\chi^{c(s)}_l$ and its extension to finite $v_F^*|\q|/\Omega$.
In Sec. \ref{sec:3} we discuss the forms of $l=1$ susceptibilities for the currents of conserved fermionic charge  and spin, and also discuss the relation between the $l=2$ charge order parameter (bilinear in fermions) and the current of a total fermionic momentum.
In Sec. \ref{sec:4} we present the results of numerical and analytical calculations to second order in the Hubbard $U$. Our key emphasis here is to understand why contributions to the $l=1$ susceptibilities from fermions at the FS and away from it are related.
In Sec. \ref{sec_new} we discuss the implications for susceptibilities of order parameters which contain an additional dependence on $k$ beyond symmetry related overall factors.
We summarize our results in Sec. \ref{sec:6}

\section{FL theory.  Diagrammatic approach}
\label{sec:2}

In this section we briefly review the diagrammatic approach to a FL and present the diagrammatic derivation of Eq. (\ref{eq:full}). We also obtain a more general expression for $\chi^{c(s)}_{l} (\q,\Omega)$ when both $\q$ and $\Omega$ are small, but the ratio $v^*_F q/\Omega$ is arbitrary. The full formula is
\begin{equation}
  \chi_{l}^{c(s)} (\q,\Omega) = \left(\Lambda^{c(s)}_l Z\right)^2  \chi^{c(s)}_{l,qp} (\q,\Omega) +\chi^{c(s)}_{l,inc} \label{eq:full_full}
\end{equation}
For arbitrary   $v^*_F q/\Omega$,  $\chi^{c(s)}_{l,qp} (\q,\Omega)$ is a complex function of all Landau parameters.
For definiteness and to make computational steps less involved, we consider two-dimensional (2D)  Galilean-invariant systems.

Our goal is to distinguish between high-energy and low-energy contributions to the susceptibility and relate $\chi^{c(s)}_{l,qp}$,  $\chi^{c(s)}_{l,inc}$, $Z$, and $\Lambda^{c(s)}_{l}$  to particular sets of diagrams. We express
different contributions to the susceptibility  via the vertex function $\Gamma^\omega (k,p)$. Here and below $k$ denotes a fermionic 3-vector, $k = (\k, \omega_k)$ and $q=(\q,\Omega)$ denotes a bosonic 3-vector. We show that $\chi^{c(s)}_{l,qp}$ is expressed via  $\Gamma^\omega (k,p)$ in which both $k$ and $p$ are on the FS,  $\chi^{c(s)}_{l,inc}$ is expressed via $\Gamma^\omega (k,p)$ in which both $k$ and $p$ are away from the FS, and $\Lambda^{c(s)}_{l}$ is expressed via $\Gamma^\omega (k,p)$ in which $k$ is on the FS and $p$ is away from it, or vice versa.  We combine our
analysis with the Landau-Pitaevskii equations \cite{AGD,Lifshitz1980} which relate the inverse quasiparticle residue $1/Z$ to $\Gamma^\omega (k,p)$ in which $k$ is on the FS and $p$ is away from it, similarly to $\Lambda^{c(s)}_{l}$. The contributions from away from the  FS are insensitive to the ratio $\Omega/|\q|$ and can be computed at $\q =0$ and $\Omega =0$. The quasiparticle part $\chi^{c(s)}_{l,qp} (q)$  depends on how the limit $\q,\Omega \to 0$ is taken.

\subsection{Perturbation theory}

The free fermion Hamiltonian is
\begin{align}
  H_{kin} &= \int dr \sum_\alpha c^\dagger_\alpha (r) \left(\frac{- \nabla^2}{2m} -\mu\right)c_\alpha (r) \nn \\
  &=   \sum_{\k \alpha} \xi_{\k}
  c^\dagger_{k, \alpha}  c_{k, \alpha}
 \label{n_1}
 \end{align}
 where $\xi_{\k} = \k^2/(2m) - \mu$,
 and we set $\hbar=1$ throughout this paper.
 The corresponding free-fermion Green's function is
 \begin{equation}
 G_0(\tk) = \frac{1}{\omega_k - \xi_{\k} + i \delta_\omega}
 \label{n_2}
 \end{equation}
 where $\delta_\omega = \delta \mbox{sgn} \omega$ and $\delta = 0^+$.
 \begin{figure}\label{fig:1}
   \includegraphics[width=3cm]{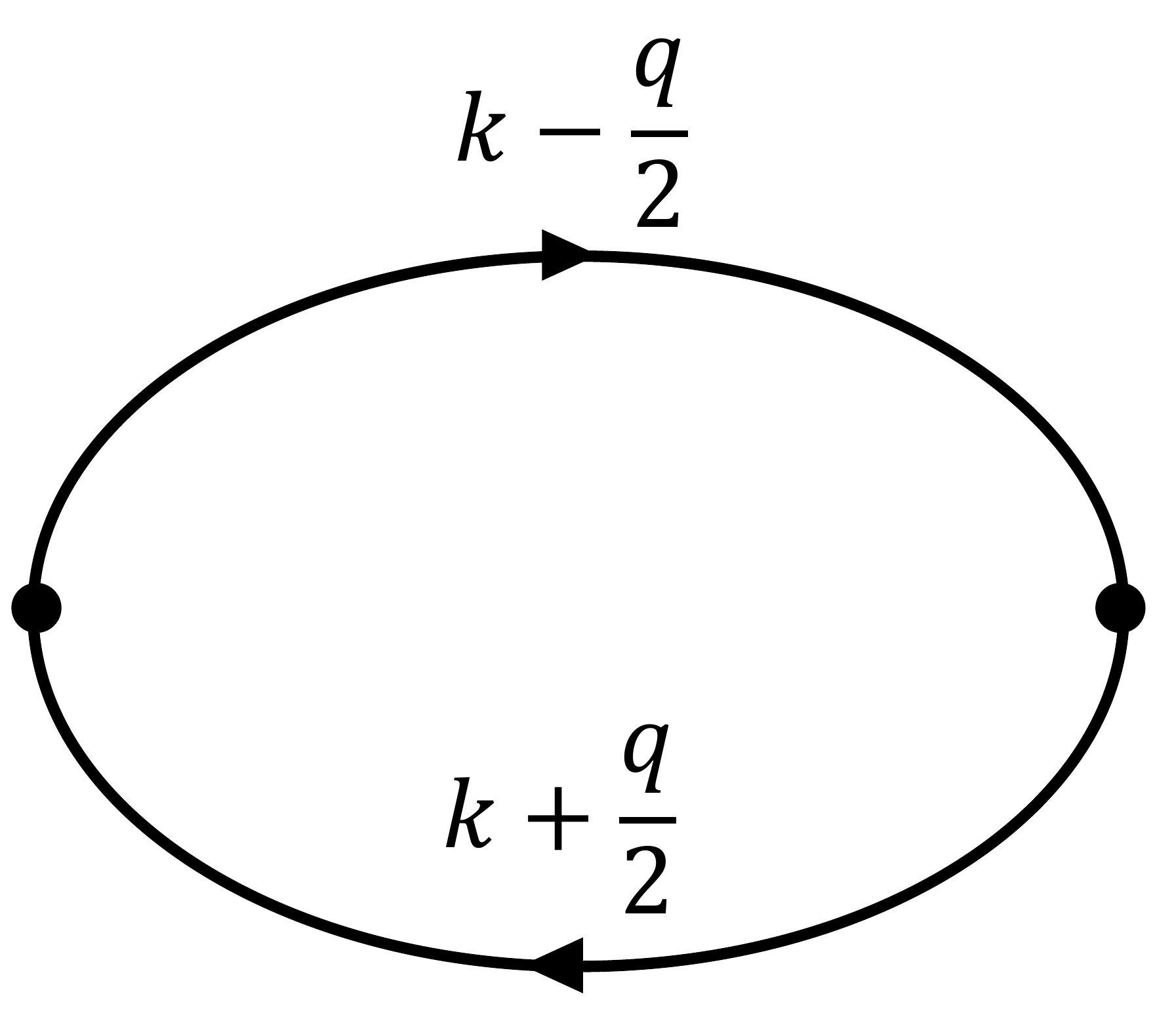}
   \caption{Free fermion susceptibility, where $\tk=(\k,\omega_k)$ and $\tq=(\q,\Omega)$. The black dots on the two sides represent form factors $\lambda_{l}(\k)$.
   }
 \end{figure}
 The free-fermion susceptibility $\chi^{c(s)}_{l,0} (\tq)$ is diagrammatically represented as the bubble made out of two fermionic propagators (Fig \ref{fig:1}) with form-factors $\lambda^{c(s)}_l$ in the vertices:
 \begin{equation}
 \chi^{c(s)}_{l,0} (\tq)=-2 \int\frac{d^3\tk}{(2\pi)^3}  \left(\lambda^{c(s)}_l (\bm{k}) \right)^2  G_0 (\tk + \frac{\tq}{2}) G_0 (\tk-\frac{\tq}{2}),
 \label{n_3_1}
\end{equation}
where the factor 2 comes from spin summation. In 2D,
\begin{equation}\label{eq:form-factor-def}
  \lambda^{c(s)}_l (\bm{k})= \cos{l\phi_{k}} |\k|^l\times f^{c(s)}_l(|\k|)
\end{equation}
where $\phi_k$ is the angle between ${\bf k}$ and $\bf q$. One may verify that the frequency integral in (\ref{n_3_1}) is non-zero only if $\xi_{\k+\q/2}$ and $\xi_{\k-\q/2}$ have opposite signs, i.e.,  it comes from the tiny range near the FS of width $O(q)$. In explicit form we have, after integrating over frequency
\begin{equation}
  \chi^{c(s)}_{l,0} (\tq)=-2\int\frac{d^2k}{(2\pi)^2}\frac{n_F(\xi_{\k-\frac{\q}{2}})-n_F(\xi_{\k+\frac{\q}{2}})}{\Omega+\xi_{\k-\frac{\q}{2}}-\xi_{\k+\frac{\q}{2}}+i\delta_{\Omega}} \left(\lambda^{c(s)}_l (\bm{k})\right)^2
 \label{n_3}
\end{equation}
where $n_F(\xi)=\Theta(-\xi)$ is a unit step function in zero temperature limit. In the case of vanishingly small $|\q|$ one can integrate over $k$ and obtain,
\begin{eqnarray}
 &&\chi_{l,0}^{\cs} (\tq)= - \frac{m}{\pi}
     \left(k_F^lf^{c(s)}_l (k_F)\right)^2
   \nonumber \\
  && \int \frac{d\phi_{k}}{2\pi} (\cos{l \phi_k})^2 \frac{ v_F |\q| \cos\phi_k}{\Omega - v_F |\q| \cos\phi_k + i \delta_\Omega}
\label{n_4}
\end{eqnarray}
In the static limit $\Omega =0, {\bf q} \to 0$ we have
 \begin{eqnarray}
 &&\chi^{c(s)}_{l=0,0}  = \frac{m}{\pi}  \left(f^{c(s)}_{l=0} (k_F)\right)^2  \nonumber \\
 && \chi^{c(s)}_{l,0} =  \frac{m}{2\pi}\left(k_F^lf^{c(s)}_{l} (k_F)\right)^2,   l>0
 \label{n_14}
 \end{eqnarray}
 We next include the interaction term
 \begin{eqnarray}
&& H_{int} = \frac{1}{2} \int d{\bf r} d{\bf r'}  \sum_{\alpha,\beta}
   c^\dagger_\alpha (r) c_\alpha (r)
   U\left(|{\bf r}-{\bf r}'|\right)
   c^\dagger_{\beta} (r') c_\beta (r') \nonumber \\
&& = \frac{1}{2V}
   \sum U (|\q|)
  c^\dagger_{\k+\q/2,\alpha}  c^\dagger_{\p-\q/2,\beta}  c_{\p+\q/2,\delta}  c_{\k-\q/2,\gamma}
  \delta_{\alpha \gamma} \delta_{\beta \delta},  \nonumber \\
  \label{n_5}
 \end{eqnarray}
 where the summation is over all momenta and all spin indices.

 \subsubsection{First order in $U(|\q|)$}
 \label{sec:first-order-uq-1}
 \begin{figure}
   \includegraphics[width=7cm]{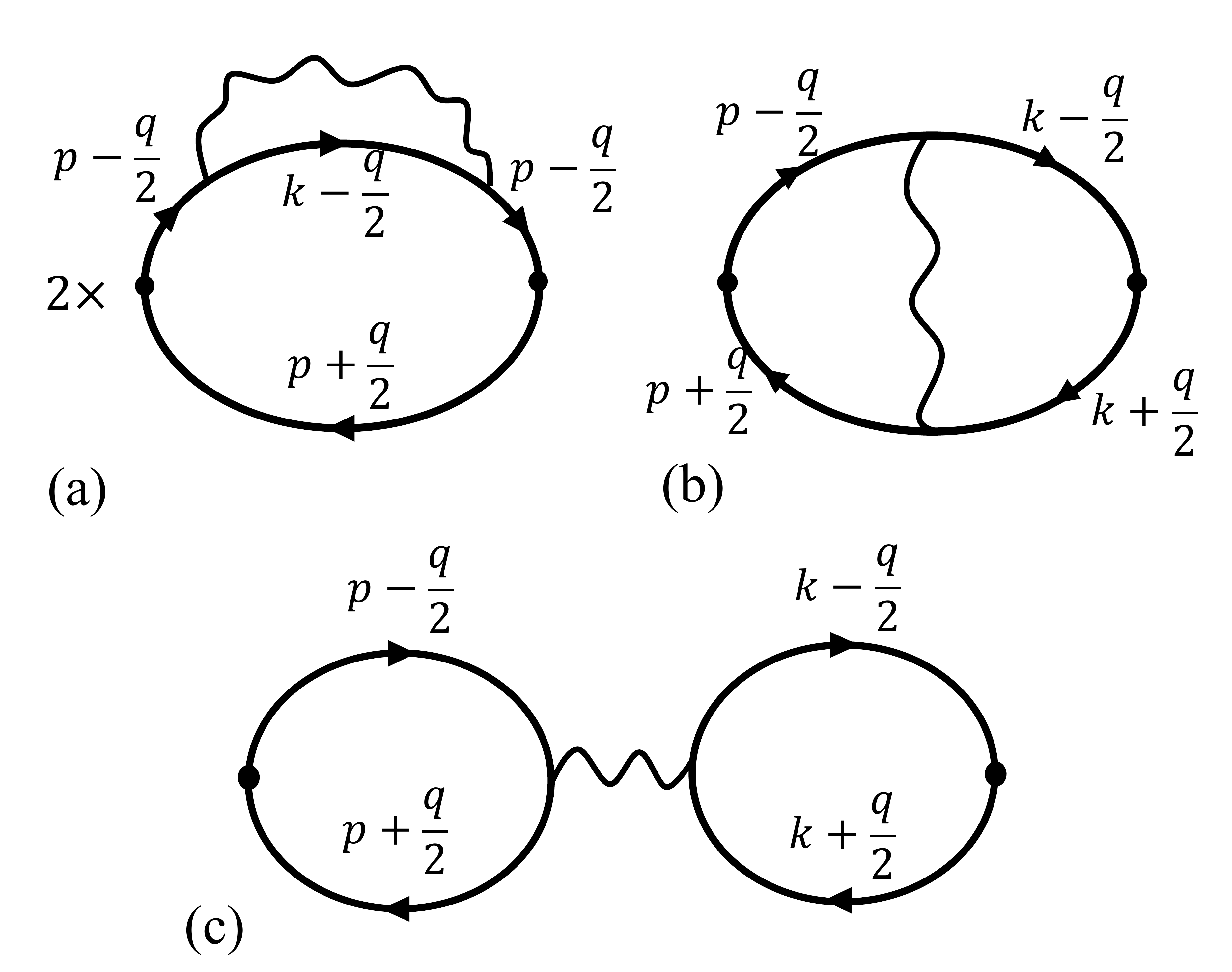}
   \caption{Corrections to $\chi_l\cs$ to first order in $U$.}
   \label{fig:chi-1st-order}
 \end{figure}
\begin{figure}
   \centering
   \includegraphics[width=7cm]{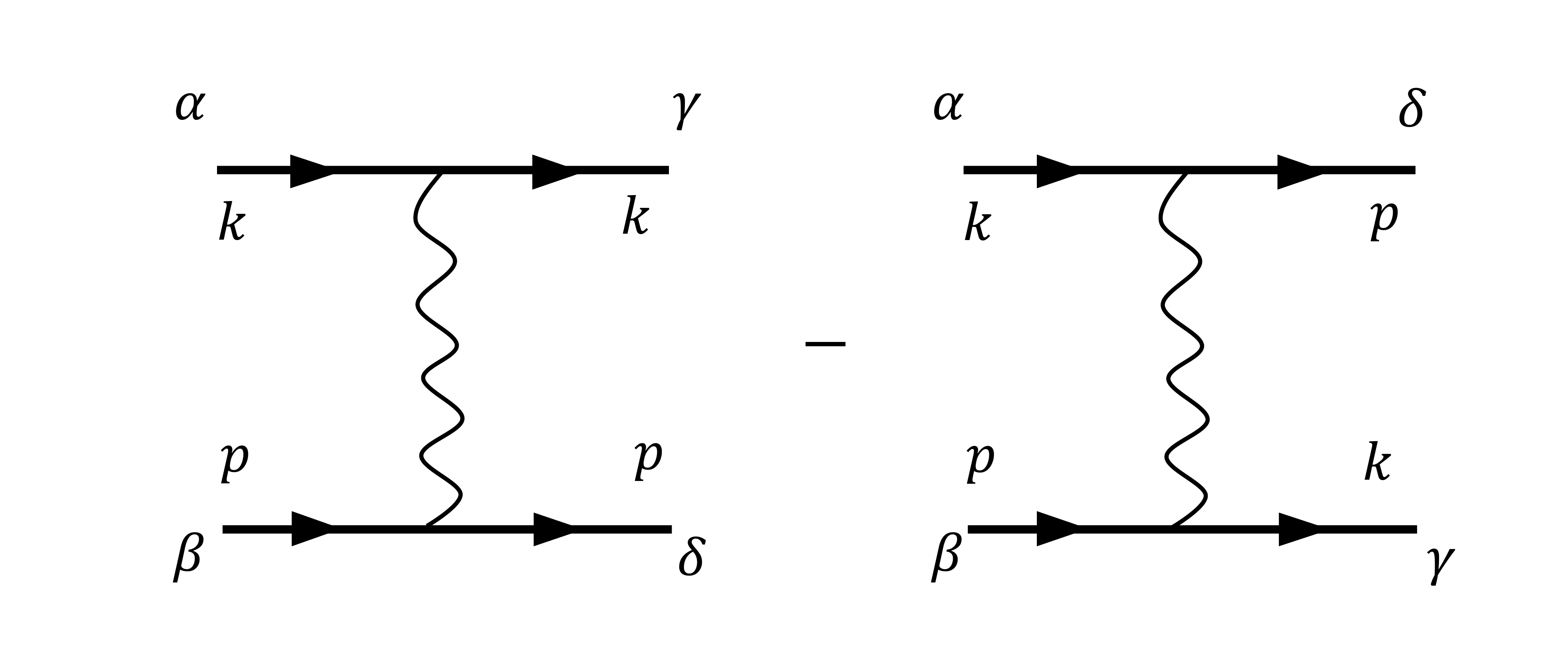}
   \caption{The vertex $\Gamma^\omega_{\alpha\beta,\gamma\delta}(\k,\k;\p,\p)$ to first order in $U$.}
   \label{fig:int-1st-order}
 \end{figure}

 To first order in  $U({\bf q})$,  there are three interaction-induced corrections to the bubble diagram for the susceptibility. They are shown in Fig. \ref{fig:chi-1st-order}.
 Diagram \ref{fig:chi-1st-order}a represents a self-energy correction. The self-energy is purely static (because $U(|\q|)$ is static) and  gives rise to mass renormalization
 $m^*/m = 1 - (1/v_F) d \Sigma/d|\k|$. One can easily verify that the integral for $\Sigma (k)$ for $\k$ near the FS is determined by $\q$ connecting points on the FS. A simple calculation yields
 \begin{equation}
   \frac{m^*}{m} = 1 - \frac{m}{2\pi} \int \frac{d \theta}{2\pi} U \left(2k_F \left|\sin\frac{\theta}{2}\right|\right) \cos{\theta}
   \label{n_6}
 \end{equation}
 and
 \begin{equation}
   \chi^{c(s)}_{l,2a} (\tq) = \left(\frac{m^*}{m}-1\right)  \chi^{c(s)}_{l,0} (\tq)
   \label{n_7}
 \end{equation}
 Diagram \ref{fig:chi-1st-order}b contains two cross-sections with internal $\tk$ and $\tp$. Because the interaction $U(\k-\p)$ is static, in each cross-section the frequency integral is again non-zero only if the dispersions have opposite signs. The result is that the integration is again confined to a narrow region near the FS. Evaluating frequency and momentum integrals, we obtain
 \begin{widetext}
   \begin{flalign}
     \chi^{c(s)}_{l,2b} (\tq) =
     \frac 1 2
     \left(\frac{m}{\pi}\right)^2
     \left(k_F^lf^{c(s)}_l (k_F)\right)^2  \int \frac{d\phi_{k}}{2\pi} \frac{d\phi_{p}}{2\pi} \cos{l \phi_k} \cos{l\phi_p} U\left(2 k_F \left|\sin
       \frac{\phi_k-\phi_p}{2}
                                \right|\right)\times\qquad\qquad\nn\\
     \frac{ v_F |\q| \cos\phi_k }{\Omega -  v_F |\q| \cos\phi_k  + i \delta_\Omega} \frac{ v_F |\q| \cos\phi_p }{\Omega -  v_F |\q| \cos\phi_p  + i \delta_\Omega}
     \label{n_8}
   \end{flalign}
 \end{widetext}
 In the static limit $\Omega=0$, $\q \to 0$,
 \bea
&& \chi^{c(s)}_{l,2b} =
\frac 1 2
\left(\frac{m}{\pi}\right)^2
        \left(k_F^lf^{c(s)}_l (k_F)\right)^2  \nonumber \\
&&\int \frac{d\phi_{k}}{2\pi} \frac{d\phi_{p}}{2\pi} \cos{l \phi_k} \cos{l\phi_p} U\left(2 k_F |\sin\frac{\phi_k-\phi_p}{2}|\right)
\eea
Finally, diagram $2c$ contains $U(0)$ and is non-zero only for charge susceptibility at $l=0$. It gives
\begin{equation}
	 \chi^{c}_{l=0,2c} (\tq) = U(0)   \left(\chi^{c}_{l=0,0} (\tq)\right)^2
   \left(f^c_{l=0} (k_F)\right)^{-2}
 \label{n_9}
 \end{equation}
 The sum of the three diagrams can be cast into a known FL form by re-expressing the results in terms of the Landau function $F_{\alpha\beta,\gamma\delta} (\k,\p) = (Z^2 m^*/\pi) \Gamma^\omega_{\alpha\beta,\gamma\delta} (k,k;p,p)$, where  $\Gamma^\omega_{\alpha\beta,\gamma\delta} (k,k;p,p)$
is the fully renormalized antisymmetrized static interaction between fermions on the FS, taken in the limit of zero momentum transfer. The antisymmetrized interaction to first order in $U$ is shown graphically in Fig. \ref{fig:int-1st-order}. To this order, $Z^2 m^*/\pi = m/\pi$.
  Combining the diagrams from this figure, we obtain
\begin{eqnarray}
 &&F_{\alpha \beta, \gamma \delta} (\k,\p) = \frac{m}{\pi} \left[U(0)\delta_{\alpha \gamma} \delta_{\beta \delta} - U(\k-\p) \delta_{\alpha \delta} \delta_{\beta \gamma}\right]  \label{n_10} \\
 &&= \frac{m}{\pi} \left[\left(U(0) - \frac{1}{2}  U(\k-\p) \right) \delta_{\alpha \gamma} \delta_{\beta \delta} - \frac{1}{2}  U(\k-\p)  {\bf \sigma}_{\alpha \gamma} {\bf \sigma}_{\beta \delta}\right] \nonumber
   \end{eqnarray}
   The two terms in the last line in (\ref{n_10}) are charge and spin components of the Landau function $F_{\alpha \beta, \gamma \delta} (\k,\p) = F^c (\k,\p) \delta_{\alpha \gamma} \delta_{\beta \delta} + F^s (\k,\p) {\bf \sigma}_{\alpha \gamma} {\bf \sigma}_{\beta \delta}$.  Each component can be further expanded in partial harmonics with different $l$ as
   \begin{equation}
 F^{c(s)} (\k,\p) = F^{c(s)}_0 + 2 \sum_{l>0} F^{c(s)}_l \cos{l \phi},
 \label{n_11}
 \end{equation}
 where $\phi = \phi_k - \phi_p$ is the angle between $\k$ and $\p$ ($|\k|=|\p| = k_F$).
 Using this expansion, one may easily check that  the sum of zero-order and first-order contributions to the static susceptibility can be cast into
 \begin{align}
   \chi^{c(s)}_l
   &= \chi_{l,0}\left(1 + F^c_{l=1} -  F^{c(s)}_{l}\right) \nn\\
   &\approx \chi_{l,0} \left(1 + F^c_{l=1}\right) \left(1 - F^{c(s)}_l\right)
   \label{n_12}
 \end{align}
 This formula is valid for all $l$, including $l=0$.
 Eq. \eqref{n_12} trivially fulfils the constraints of
 Eqs \eqref{n2} and \eqref{n3} for the simple reason that
 to this order,
 $F_l^c = F_l^s$ for all $l > 0$.
\subsubsection{Higher orders in $U(q)$, static limit}
\label{sec:higher-orders-uq}
We now move to higher orders in $U$, still considering the static limit $\Omega =0, \q \to 0$.  Within RPA, higher-order diagrams are treated as series of ladder graphs ($l>0$) or ladder and bubble graphs ($l=0$),   Each element of the ladder/bubble series contains the product of two fermionic Green's functions, dressed by static self-energy.  The two Green's functions have the same frequency and their momenta differ by ${\bf q}$. Within this approximation, a non-zero contribution to susceptibility from each cross-section comes from the states very near the FS, where the poles in the two fermionic Green's functions, viewed as functions of frequency,  are shifted in different directions from the real frequency axis.  A simple analysis shows that the series is geometric and its sum yields
\begin{equation}
\chi^{c(s)}_{l, RPA} = \chi_{l,0}  \frac{1 + F^c_{l=1}}{1 + F^{c(s)}_l}
\label{n_16}
\end{equation}
The RPA susceptibility obviously diverges when $F^{c(s)}_l = -1$,
except for the special case of $F_l\cs = F_1^c$, as occurs e.g. for $l=1$ if we require that the interaction is purely static, see the previous section and our comments in the Introduction.

We next go beyond RPA.  A diagram for $\chi_l\cs$ at any loop order is represented by a series of ladder segments separated by interactions. In each of these ladders there is an integration over both high-energy and low-energy frequencies and momenta. To obtain $\chi^{c(s)}_{l}$, we follow earlier diagrammatic studies \cite{Eliashberg1962,Finkelstein2010,Chubukov2014a} and and re-arrange perturbation series by assembling contributions to $\chi^{c(s)}_{l}$ from diagrams with a given number $M$ of ladder segments with poles shifted into different directions from the real frequency axis, and then sum up contributions from the sub-sets with different $M =0,1,2$, etc.

\begin{figure}
  \centering
  \includegraphics[width=0.3\hsize]{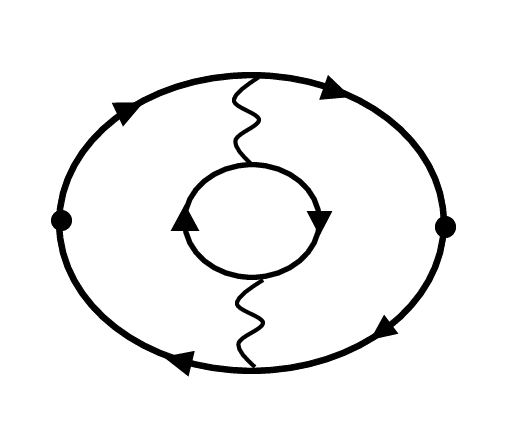}
  \caption{Example of a higher order contribution to $\chi_l\cs$. At this order, the static interaction acquires dynamics due to particle-hole screening. The diagram's computation is split into three (see Sec. \ref{sec:higher-orders-uq}). It belongs to the $M=0$ sector when both bubbles are evaluated away from the FS, to $M=1$ when one is evaluated on the FS and one away from it, and to $M=2$ when both are evaluated at the FS.}\label{fig:bubble-screening}
\end{figure}

We start with $M=0$.  The corresponding  contributions to the susceptibility contain products of $G^2 (\k, \omega_k)$. Taken alone, each such term will vanish after integration over frequency. The total $M=0$ contribution then vanishes to first order in $U(\q)$ because the static interaction does not affect the frequency integration. However, at second  and higher orders in $U(\q)$, the interaction gets screened by particle-hole bubbles and becomes a dynamical one. An example of second-order susceptibility diagram  with screened interaction inserted into the bubble is shown in Fig. \ref{fig:bubble-screening}.
This screened dynamical interaction contains a Landau damping term, which is non-analytic in both half-planes of complex frequency. As a result, the product of $G^2 (\k, \omega_k)$ and the dressed interaction at order $U^2$ and higher has both a double pole and a branch cut. A pole can be avoided by closing the integration contour in the appropriate frequency half-plane,
but the branch cut is unavoidable, and its presence renders the frequency integral finite.
Since there is no splitting, relevant fermionic $\omega_k$ and ${\bf k}$ are not confined to the FS and are generally of order $E_F$ (or bandwidth). Fermions at such high energies have a finite damping, i.e., are not fully coherent quasiparticles. By this reason, the $M=0$ contribution to $\chi^{c(s)}_{l}$ is labeled as an incoherent one, $ \chi^{c(s)}_{l, M=0} = \chi^{c(s)}_{l,inc}$ (although at small $U$ fermions with energies of order $E_F$ are still mostly coherent).
\begin{figure}
  \centering
  \includegraphics[width=\hsize]{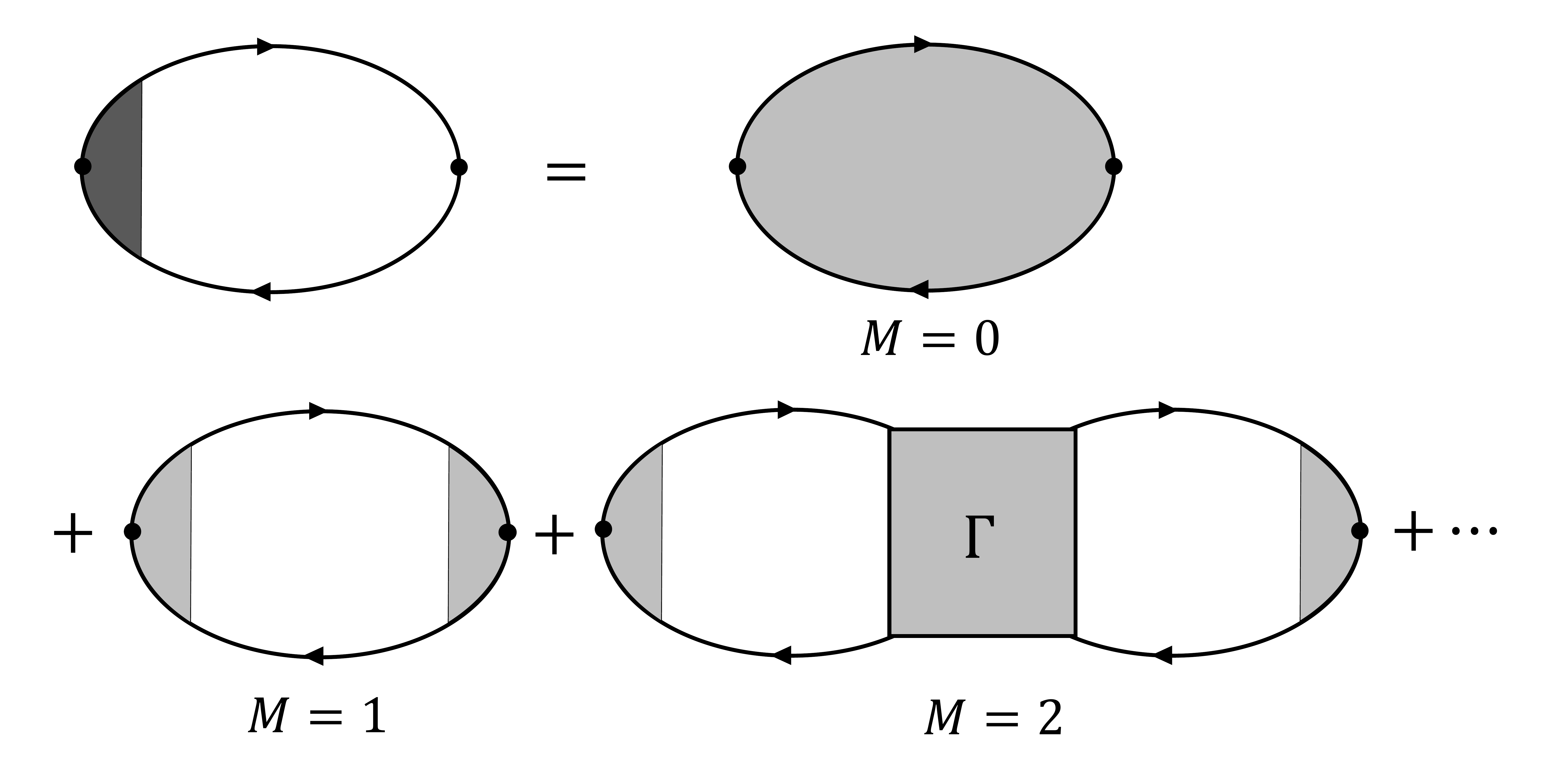}
  \caption{The ladder series of diagrams for the static susceptibility $\chi_l\cs$. The exact $\chi_l$ is represented as a series  $M=0,1,2,\ldots$ of bubbles comprised of Green's functions with poles on opposite halves of the complex frequency plane, i.e. whose contributions are computed close to the FS.}
  \label{fig:bubble-series}
\end{figure}

We next move to the $M=1$ sector.  Here we select the subset of diagrams with one cross-section, in which we pick up the contribution from $G({\bf k}, \omega_k) G({\bf k}+{\bf q}, \omega_k)$ from the range where the poles in the two Green's functions are in different half-planes of complex frequency.  The sum of such diagrams can be graphically represented by the skeleton diagram in Fig. \ref{fig:bubble-series} labeled $M=1$. The internal part of this diagram gives $Z^2 (m^*/m) \chi_{l,0} (\tq)$, where $\chi_{l,0} (\tq)$ is given by (\ref{n_4}).
The side vertices contain $\Lambda_1\cs\lambda_l\cs(k_F)$, i.e. the product of the bare form-factor (which we already incorporated into $\chi_{l,0} (\tq)$), and the contributions from all other cross-sections, in which $G({\bf k}, \omega_k) G({\bf k}+{\bf q}, \omega_k)$ is approximated by $G^2({\bf k}, \omega_k)$.  These contributions  would vanish if we used a static $U(|\q|)$ for the interaction, but again become non-zero once we include dynamical screening  at order $U^2$ and higher. Similarly to the $M=0$ sector, the difference  $\Lambda_l^{c(s)} -1$ is determined by fermions with energies of order $E_F$. Note, however, that in the $M=0$ sector, all internal energies are of order $E_F$. In the $M=1$ sector, internal energies for the vertices $\Lambda_l^{c(s)}$ are of order $E_F$, but external $\omega_k$ are infinitesimally small, and external ${\bf k}$ are on the FS.  Overall, the contribution to the static susceptibility from the $M=1$ sector is
\begin{equation}
  \chi^{c(s)}_{l, M=1} = \left(Z \Lambda^{c(s)}_l\right)^2 \frac{m^*}{m} \chi_{l,0}\cs
  \label{n5}
\end{equation}
Sectors with $M=2$, $M=3$ are  the subsets of diagrams with $2, 3,\ldots$ cross-sections in which we split the poles of the Green's functions with equal frequencies and momenta separated by ${\bf q}$.  In the cross-sections in between the selected ones $G({\bf k}, \omega_k) G({\bf k}+{\bf q}, \omega_k)$  is again approximated by $G^2({\bf k}, \omega_k)$.  The contribution from the $M=2$ sector is represented by the skeleton diagram in Fig. \ref{fig:bubble-series} labeled $M=2$. It contains fully dressed side vertices $\Lambda^{c(s)}_l$ and a fully dressed anti-symmetrized static interaction between fermions on the FS.  One can easily verify that this interaction appears with the prefactor $Z^2 (m^*/m)$, i.e.,  the extra factor in the $M =2$ sector compared to $M=1$ is the product of $\chi_{l,0}$ and the corresponding component of the Landau function.  Using (\ref{n_11}) we then obtain
\begin{equation}
 \chi^{c(s)}_{l, M=1}+
  \chi^{c(s)}_{l, M=2} = \left(Z \Lambda^{c(s)}_l\right)^2 \frac{m^*}{m} \chi_{l,0}\cs \left(1 - F^{c(s)}_l\right)
  \label{n6}
\end{equation}
(the minus sign comes from the number of fermion bubbles.) A simple bookkeeping analysis
shows that contributions from sectors with larger $M$ form a geometric
 series, which transform $1 - F^{c(s)}_l$ into $1/(1 + F^{c(s)}_l)$.  Collecting all contributions, we reproduce Eq. (\ref{eq:full}).

\subsubsection{The susceptibility $\chi^{c(s)}_l ({\bf q}, \Omega)$ at finite $\Omega/v^*_F|{\bf q}|$.}
\label{sec:susc-chics_l-bf}

We now extend the analysis to the case when both transferred momentum ${\bf q}$ and transferred frequency $\Omega$ are  vanishingly small, but the ratio $\Omega/v^*_F |{\bf q}|$ is finite.  The computational steps are the same as for static susceptibility. The contribution to $\chi^{c(s)}_l (\tq)$ from the $M=0$ sector and the vertex function $\Lambda^{c(s)}_l$ do not depend on the ratio of $\Omega/(v^*_F |{\bf q}|)$ and remain the same as in the static case. However, the integrand in  the expression for $\chi_{l,0} (\tq)$, Eq. (\ref{n_4}),  now contains a non-trivial angular dependence via
$ v_F|\q|\cos\phi_k/(\Omega - v_F|\q|\cos\phi_k + i \delta_\Omega)$.  This makes the computation of series with $M=1,2, \ldots$ more involved.

Consider first the limit $\Omega \ll v_F |\q|$. For even $l$, the free-fermion susceptibility is
\begin{eqnarray}
  && \chi^{c(s)}_{l,0}(q) =
     \frac{m}{\alpha_l\pi} \left(k^l_F  f^{c(s)}_l (k_F)\right)^2 \left(1 + \alpha_l \frac{i\Omega}{v_F |\q|} \right) \nonumber \\
  && =
     \chi^{c(s)}_{l,0}\left(1 +
     \alpha_l
     \frac{i\Omega}{v_F |\q|} \right)
\end{eqnarray}
where $\alpha_l=1$ if $l=0$ and $\alpha_l=2$ if $l = 2 m$, $m >0$.
For odd $l$, the expansion in $\Omega$ starts with $\Omega^2$. The total contribution from the $M=1$ sector still is proportional to $\chi_{l,0}$:
\begin{eqnarray}
  && \chi^{c(s)}_{l,M =1} (\tq) \approx
  \frac{m^*}{\alpha_l\pi} \left(Z \Lambda^{c(s)}_l\right)^2  \left(k^l_F  f^{c(s)}_l (k_F)\right)^2 \nonumber \\
   && \left(1 + \frac{m^*}{m} \alpha_l\frac{i \Omega}{v_F |{\bf q}|} \right) \nonumber \\
&& =  \left(Z \Lambda^{c(s)}_l\right)^2 \frac{m^*}{m}\chi^{c(s)}_{l,0}\left(1 + \frac{m^*}{m}
 \alpha_l
\frac{i \Omega}{v_F |{\bf q}|}\right)
\label{n12}
\end{eqnarray}
In the contribution from the $M=2$ sector, the $i\Omega/v^*_F |{\bf q}|$ term can be taken from the cross-section on the right or on the left. This gives a combinatoric factor of $2$. Then
\begin{eqnarray}
&& \chi^{c(s)}_{l,M =1}(q)+\chi^{c(s)}_{l,M =2} (\tq) \approx
    \left(Z \Lambda^{c(s)}_l\right)^2  \frac{m^*}{m} \chi_{l,0}\cs \nonumber \\
&& \left(1 - F^{c(s)}_l  + (1 - 2F^{c(s)}_l) \frac{m^*}{m}\alpha_l\frac{i \Omega}{v_F |{\bf q}|}
\right)
 \label{n14}
\end{eqnarray}
For the contribution from the $M=3$ sector the same reasoning yields the combinatoric factor of 3 and so on. Using
\begin{equation}
1 - 2F^{c(s)}_l + 3 \left(F^{c(s)}_l\right) + ... = \frac{1}{(1 + F^{c(s)}_l)^2}
\end{equation}
we obtain
\begin{equation}
  \chi^{c(s)}_{l} (\tq)  = \left(Z \Lambda^{c(s)}_l\right)^2   \chi^{c(s)}_{l,qp} (\tq) + \chi^{c(s)}_{l,inc}
  \label{n16}
\end{equation}
where to order $\Omega/|\q|$, for even $l$,
\begin{equation}
  \chi^{c(s)}_{l,qp} (\tq) =
   \chi_{l,0}\cs
   \left(\frac{m^*/m}{(1 + F^{c(s)}_l)} +
    \alpha_l
   \frac{i \Omega}{v_F |{\bf q}|}
      \left(\frac{m^*/m}{1 + F^{c(s)}_l} \right)^2 \right)
      \label{n15}
\end{equation}
For $l=0$ this result has been obtained before \cite{Nozieres1999}.

In the opposite limit $\Omega \gg v^*_F |{\bf q}|$ we have
\begin{align}
  \chi_{l,0} (\tq) \approx - \frac{m}{\pi}
  \left(\frac{v_F |{\bf q}|}{\Omega}\right)^2 \left(k^l_F f_l (k_F)\right)^2
  \nn\\
  \int \frac{d\phi_{k}}{2\pi} (\cos{l \phi_k})^2  (\cos{\phi_k})^2
  \label{n7}
\end{align}
The presence of $|\q|^2/\Omega^2$ in the susceptibility for $l=0$ is a natural consequence of the fact that the total fermionic charge and spin are conserved quantities, i.e., they don't change when we probe the system at different times. For free fermions, this holds for all $l$ because
all partial fermionic densities at a given direction of ${\bf k}$ are separately conserved, hence  $\chi_{l,0} ({\bf q} =0, \Omega)$ must vanish for an any angle-dependent form-factor. The contribution from the $M=1$ sector is,
\begin{flalign}
  \chi^{c(s)}_{l,M =1} (\tq) \approx -& \frac{m}{\pi}
   \left(Z \Lambda^{c(s)}_l\right)^2
  \frac{m}{m^*}  \left(\frac{v_F |{\bf q}|}{\Omega}\right)^2 \left(k^l_F f_l (k_F)\right)^2 \times\quad \nonumber \\
  &\int \frac{d\phi_{k}}{2\pi} (\cos{l \phi_k})^2  (\cos{\phi_k})^2.
  \label{n8}
\end{flalign}
The  overall $m/m^*$ factor is due to one $m^*/m$ factor from the integration over momentum and an $(m/m^*)^2$ from the expansion to second order in  $v^*_F |\q|/\Omega$. From the $M=2$ sector we have, at order $|\q|^2/\Omega^2$
\begin{widetext}
  \begin{equation}
    \chi^{c(s)}_{l, M=2} = -
    \left(\frac{v_F |{\bf q}|}{\Omega}\right)^2 \left(\frac{m}{\pi}\right)
   \left(Z \Lambda^{c(s)}_l\right)^2 \frac{m}{m^*} \left(k^l_F f^{c(s)}_l (k_F)\right)^2
    \int \int \frac{d \phi_k}{2\pi} \frac{d \phi_p}{2\pi} (\cos{l \phi_k}) (\cos{l \phi_p}) (\cos{\phi_k}) (\cos{\phi_p})
    F^{c(s)} (\phi_k - \phi_p)
    \label{n9}
  \end{equation}
  Substituting $F^{c(s)}$ from  Eq. (\ref{n_11}), we obtain
  \begin{eqnarray}
    &&\chi^{c(s)}_{l=0, M=2} = -\frac{1}{2}\left(\frac{v_F |{\bf q}|}{\Omega}\right)^2 \left(\frac{m}{\pi}\right) \left(Z \Lambda^{c(s)}_l\right)^2  \left(f^{c(s)}_{l=0} (k_F)\right)^2 \frac{m}{m^*}
   F^{c(s)}_1 \nonumber \\
    &&\chi^{c(s)}_{l=1, M=2} =
       -\frac 1 8
       \left(\frac{v_F |{\bf q}|}{\Omega}\right)^2 \left(\frac{m}{\pi}\right)
   \left(Z \Lambda^{c(s)}_l\right)^2  \left(k_F f^{c(s)}_{l=1} (k_F)\right)^2 \frac{m}{m^*}
       \left(2 F_0\cs + F_2\cs\right) \nonumber \\
    &&\chi^{c(s)}_{l>1, M=2} = -\frac{1}{8} \left(\frac{v_F |{\bf q}|}{\Omega}\right)^2 \left(\frac{m}{\pi}\right)
   \left(Z \Lambda^{c(s)}_l\right)^2  \left(k^l_F f^{c(s)}_{l} (k_F)\right)^2 \frac{m}{m^*} \left( F^{c(s)}_{l-1} + F^{c(s)}_{l+1}\right)  \nonumber \\
    \label{n10}
  \end{eqnarray}
  The contribution from the sectors with $M >2$ contains higher power of $|\q|/\Omega$.  Hence, to order $|\q|^2/\Omega^2$, the full result for the dynamical susceptibility is
  \begin{eqnarray}
    &&\chi^{c(s)}_{l=0} (\tq)  =
    -\frac{1}{2}
     \left(\frac{v_F |{\bf q}|}{\Omega}\right)^2 \chi^{c(s)}_{l=0,0} \left(Z \Lambda^{c(s)}_l\right)^2 \frac{m}{m^*} \left(1 +  F^{c(s)}_1\right)
    + \chi^{c(s)}_{l=0,inc}
     \nonumber \\
    &&\chi^{c(s)}_{l=1} (\tq)  =
     -\frac{3}{4}
     \left(\frac{v_F |{\bf q}|}{\Omega}\right)^2 \chi^{c(s)}_{l=1,0} \left(Z \Lambda^{c(s)}_l\right)^2 \frac{m}{m^*} \left(1 +
       \frac{2}{3} F^{c(s)}_0 + \frac{1}{3} F^{c(s)}_2\right)
        + \chi^{c(s)}_{l=1,inc}
         \nonumber \\
    &&\chi^{c(s)}_{l>1} (\tq)  =
    -\frac{1}{2}
    \left(\frac{v_F |{\bf q}|}{\Omega}\right)^2 \chi^{c(s)}_{l,0} (\tq) \left(Z \Lambda^{c(s)}_l\right)^2 \frac{m}{m^*} \left(1 + \frac{1}{2} \left( F^{c(s)}_{l-1} + F^{c(s)}_{l+1}\right)\right)
    +  \chi^{c(s)}_{l,inc}
     \nonumber \\
    \label{n11}
  \end{eqnarray}
\end{widetext}
For $l=0$ this result has been obtained in Ref. \onlinecite{Leggett1965}.

For a generic $\Omega/v_F |{\bf q}|$,  the full expression for $\chi^{c(s)}_l (\tq)$ is rather involved for all $l$, including $l=0$.  As an illustration,  consider the seemingly simplest case $l =0$
and set $f_0 (|k|) =1$  (i.e., consider susceptibilities for spin and charge order parameters).  Due to spin/charge conservation $Z \Lambda^{c(s)}_{l=0} =1$ and $\chi^{c(s)}_{l=0,inc} =0$, so $\chi^{c(s)}_{l=0} (\tq) = \chi^{c(s)}_{l=0,qp} (\tq)$.

The full dynamical $\chi^{c(s)}_{l=0,qp} (\tq)$ is given by series of bubbles, each is determined by fermions in the vicinity of the FS.  The integration over frequency and over fermionic dispersion can be performed independently in each bubble, but angular integration is, in general, rather involved, because the interaction between the bubbles with internal momenta ${\bf k}$ and ${\bf p}$ is expressed via the Landau function
$ F^{c(s)} (\k,\p)$, Eq. (\ref{n_11}), and the latter dependens on $\phi = \phi_k - \phi_p$.   It is sufficient to analyze the first few orders in the expansion in powers of $ F^{c(s)} (\k,\p)$ to  understand that the full result is
\begin{equation}
  \chi^{c(s)}_{l=0,qp} (\tq) = \frac{m^*}{\pi}  \frac{{\bar \chi} (\tq)}{1 + F^{c(s)}_{l=0} {\bar \chi} (\tq)}
  \label{c1}
\end{equation}
where ${\bar \chi} (\tq)$ is given by series of terms
\begin{widetext}
  \begin{eqnarray}
    {\bar \chi} (\tq) &=& K_0 - 2 \sum_{n,m >0} F^{c(s)}_{n} K_n K_m \times \nonumber \\
                      && \left[\delta_{n,m} -  \sum_{m_1 >0} Q_{n,m_1} F^{c(s)}_{m_1}\left[\delta_{m_1,m} -  \sum_{m_2 >0} Q_{m_1,m_2} F^{c(s)}_{m_2}
                         \left(\delta_{m_2,m} - ...\right) \right] \right]
                         \label{c2}
  \end{eqnarray}
\end{widetext}
where $\delta_{n,m}$ is  Kroneker  symbol and
\begin{subequations}\label{c3}
  \begin{equation}
    \label{eq:Qdef}
    Q_{n,m} =  K_{n+m} + K_{n-m}.
  \end{equation}
  Here
  \begin{align}
    K_n (\tq) &= -\int \frac{d \theta}{2\pi} \cos{n \theta} \frac{v^*_F |\q| \cos {\theta}}{\Omega - v^*_F |\q| \cos{\theta} + i \delta_{\Omega}} \nonumber \\
              &= \delta_{n,0} - \frac{\alpha}{\sqrt{\alpha^2-1+i\delta}}(\alpha-\sqrt{\alpha^2-1})^{|n|},
  \end{align}
  and $\alpha = \Omega/v_F^*|\q|$.
\end{subequations}
In explicit form
\begin{eqnarray}
  && K_0 (\tq) = 1 - \frac{\Omega}{\sqrt{\Omega^2 - (v^*_F |\q|)^2 + i \delta}} \nonumber \\
  && K_1 (\tq)= \frac{\Omega}{v^*_F |\q|} \left( 1- \frac{\Omega}{\sqrt{\Omega^2 - (v^*_F |\q|)^2 + i \delta}}\right)\nonumber \\
  && K_2 (\tq) =2 \left(\frac{\Omega}{v^*_F |\q|}\right)^2  \nonumber \\
  &&
       + \frac{\Omega}{\sqrt{\Omega^2 - (v^*_F |\q|)^2 + i \delta}} \left(1 - 2\left(\frac{\Omega}{v^*_F |\q|}\right)^2\right)
     \label{c8}
\end{eqnarray}

Eq. (\ref{c2}) can be equivalently re-expressed as
\begin{equation}
  {\bar \chi} (\tq) = K_0 - 2 \sum_{n,m >0} F^{c(s)}_{n} K_n K_m S^m_n
  \label{c4}
\end{equation}
where $S^m_n$ is the solution of the matrix equation
\begin{equation}
  S^m_n + \sum_{m_1 >0} Q_{n,m_1} F^{c(s)}_{m_1} S^m_{m_1} = \delta_{n,m}
  \label{c5}
\end{equation}
In the static limit $K_0 =1$, $K_{n>0} =0$. Then ${\bar \chi} (\tq) =1$, and Eq. (\ref{c1}) reduces to Eq. (\ref{n1}) for the static susceptibility.
For a generic $\Omega/v^*_F |{\bf q}|$ a closed-form expression for $\chi^{c(s)}_{l=0,qp} (\tq)$ can be obtained if only a few Landau parameters are sizable, e.g., if we assume that
 $|F_l| \ll |F_0|,|F_1|$ for all $l > 1$.
In this situation, only one term in each sum in
(\ref{c4}) and (\ref{c5}) survives, and these two equations simplify to
\begin{equation}
  {\bar \chi} (\tq) = K_0 - 2  F^{c(s)}_{1} K^2_1 S^1_1
  \label{c6}
\end{equation}
and
\begin{equation}
  S^1_1 \left(1  + Q_{1,1} \Gamma^{c(s)}_{1}\right)  = 1
  \label{c7}
\end{equation}
Using $Q_{1,1} = K_0 + K_2$  we find $S^1_1 = 1/(1 + ( K_0 + K_2) F^{c(s)}_{1})$.  Substituting this into (\ref{c6}) and then substituting (\ref{c6}) into (\ref{c1}), we obtain
\begin{equation}
  \label{n18}
  \chi^{c(s)}_{l=0,qp} (\tq) = \frac{m^*}{\pi} \frac{K_0  - \frac{2 F^{c(s)}_1 K^2_1}{1  +  F^{c(s)}_1 \left(K_0 + K_2 \right)}}{1 + F^{c(s)}_0 K_0 - \frac{2 F^{c(s)}_0 F^{c(s)}_1 K^2_1}{1  +  F^{c(s)}_1 \left(K_0+ K_2\right)}}
\end{equation}
The same result has been obtained
previously \cite{MaslovBoltzmann}
using a Boltzmann equation approach.
At $\Omega/v_F^* |{\bf q}| \gg 1$, we have $K_0 (\tq) \approx -(1/2) (v_F |{\bf q}|/\Omega)^2$,  $K^2_1 (\tq) \approx (1/4) (v_F |{\bf q}|/\Omega)^2$,  $K_2 (\tq) \approx -(3/8) (v_F |{\bf q}|/\Omega)^2$.  Substituting into (\ref{n18}) we obtain $\chi^{c(s)}_{l=0} (\tq) = -(1/2) (v_F |{\bf q}|/\Omega)^2 (1 +  F^{c(s)}_1)$, as in Eq. (\ref{n11}).

\section{Susceptibilities of the currents of conserved order parameters}
\label{sec:3}

In this section we discuss the relationship between order parameters associated with conserved ``charges'' (to be distinguished from the specific electric charge) and their currents. We review the derivation of the continuity equation for susceptibilities of these order parameters (Refs. \onlinecite{Leggett1965,Kiselev2017} and show that this equation explicitly connects high energy properties of a FL, namely $\chi_{l,inc}\cs,\Lambda_l\cs,Z$, with low-energy properties, namely $\chi_{l,qp}$. We discuss the implications for the $l=0,1$ channels and obtain Eqs. \eqref{n2}-\eqref{n3}. Finally we discuss the implications of the continuity equation for the $l\geq 1,2$ channels. Our focus here
 is to identify
the constraints placed by the conservation law on high- and low- energy FL properties. We will then analyze these constraints microscopically in Sec. \ref{sec:4}.

\subsection{The continuity equation for charge and current susceptibilities}
\label{sec:cont-equat-susc}

A conserved ``charge'' is an operator $\hat \rho(\q,t)$ that commutes with the Hamiltonian at $\q = 0$, so that it does not evolve in the Heisenberg picture,
\begin{equation}
  \label{eq:conservation-law}
  \frac{\partial\r}{\partial t} = \frac{1}{i}[\r,H] = 0.
\end{equation}
Examples of such charges are the number (or electric charge) and spin density in the model of Sec. \ref{sec:2}:
$\r^{c}_{l=0}$ and ${\bf \r}^s_{l=0}$ from Eq. (\ref{eq:rho-def}) with constant form-factors. The continuity equation for a conserved charge $\r$ can be derived in the Heisenberg picture:
\begin{equation}
  \label{eq:drho-dt}
  \frac{\partial\r(\q,t)}{\partial t} = \frac{1}{i} [\r(\q,t), H=H_{kin}+H_{int}] \equiv -
  i
  \q \cdot \J.
\end{equation}
The continuity equation relates the susceptibilities of order parameters associated with $\hat\rho$ and $\hat\J$,
\begin{align}
  \label{eq:susc-defs}
  \chi_\rho &= \langle[\r(\q,t),\r(-\q,t')]\rangle \\
  \chi_J &= \langle[(\J)^{i}(\q,t),(\J)^{j}(-\q,t')]\rangle
\end{align}
Taking the derivative $\partial_t\partial_{t'}\chi_\rho$ and transforming to the frequency domain we obtain
\begin{equation}
  \label{eq:cont-susc}
  \Omega^2 \chi_\rho(\tq) =
  \sum_{m,n}q_m \left[\chi^{mn}_J(\tq) -\chi^{mn}_J(\q,0) \right] q_n.
\end{equation}
Here, the sum is over spatial indices $m,n=\{x,y\}$. Equivalently we may write,
\begin{equation}
  \label{eq:cont-susc-par}
  (\Omega/q)^2\chi_\rho(\tq) =
   \chi_J^\parallel(\tq) - \chi_J^\parallel(\q,0).
\end{equation}
Here we have defined the longitudinal component of the susceptibility $\hat q \cdot \chi_J \cdot \hat q$. Note, that the RHS of Eqs. \eqref{eq:cont-susc}+\eqref{eq:cont-susc-par} includes only the time dependent part of $\chi_J$. This is an automatic consequence of taking the time derivative of $\chi_\rho$ and going to the Fourier domain.

Let's  assume that both $\r$ and $J$ are expressed via bilinear combinations of fermions with some given $l$. We then can use Eq. \eqref{eq:full_full} and write
\begin{align}
  \label{eq:full-rho}
  \chi_\rho (q,\Omega) = (\Lambda_\rho Z)^2\chi_{\rho,qp} (q,\Omega) + \chi_{\rho,inc},\\
   \chi_J (q, \Omega) = (\Lambda_J Z)^2\chi_{J,qp} (q,\Omega) + \chi_{J,inc},  \label{eq:full-j}
\end{align}
 Combining these expressions and Eq. ~\eqref{eq:cont-susc-par} we express the current susceptibility via the susceptibility of a conserved charge.

\subsection{Implication of conservation laws for the susceptibilities}
\label{sec:impl-cons-laws}

For a conserved charge ~\eqref{eq:cont-susc-par} yields
 \begin{equation}
  \label{eq:chi-rho-vanish}
  \chi_\rho(\q=0,\Omega) = 0.
\end{equation}
We also  we recall that the coherent part of $\chi_\rho$, which corresponds to the $M=1,2,\ldots$ diagrams of Fig. \ref{fig:bubble-series}, vanishes at $\q = 0$. Thus, Eq. \eqref{eq:chi-rho-vanish} also implies
\begin{equation}
  \label{eq:chi-rho-inc-zero}
  \chi_{\rho,inc} = 0.
\end{equation}
Finally, the relation $\Lambda_\rho Z =1$ follows from the fact that $\Lambda_\rho$ and $1/Z$ are identically expressed via the vertex $\Gamma^\omega$,
\begin{figure}
  \includegraphics[width=7cm]{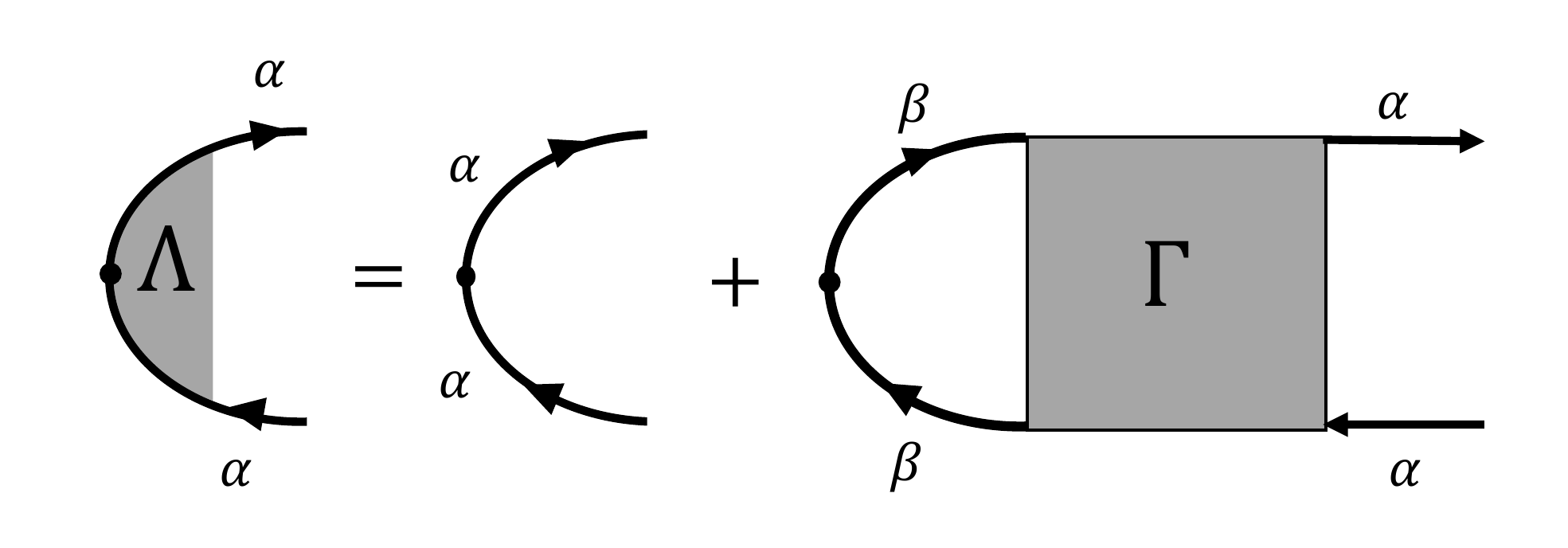}
  \caption{Relation between a 3-leg vertex $\Lambda$ and a 4-leg vertex $\Gamma$, for a conserved charge density.}\label{fig:3-4vertex}
\end{figure}
\begin{subequations}
  \begin{align}
    \label{eq:xxxx1}
    &\Lambda_\rho = 1  - \frac{i}{2 k_F}\sum_{\alpha\beta}\int \frac{d^3 k}{(2\pi)^3} \Gamma^\omega_{\alpha\beta,\alpha\beta}(k_F\hat p, \k)(G_k^2)^\omega \frac{\lambda_\rho (k)}{\lambda_\rho (k_F)} \\
    &   \frac{1}{Z} = 1  - \frac{i}{2 k_F}\sum_{\alpha\beta}\int \frac{d^3 k}{(2\pi)^3} \Gamma^\omega_{\alpha\beta,\alpha\beta}(k_F\hat p, \k)(G_k^2)^\omega \frac{\lambda_\rho (k)}{\lambda_\rho (k_F)}
      \label{eq:xxxx2}
  \end{align}
\end{subequations}
where  $(G_q^2)^\omega = \mbox{lim}_{\Omega\to 0} G(\q,\omega)G(\q,\omega+\Omega)
= G^2(\q,\omega)$ is the regular part of the product of two Green's functions, For the vertex, Eq. (\ref{eq:xxxx1})  follows from Fig. \ref{fig:3-4vertex} (and is valid for a conserved "charge" in  both charge and spin channels, while
for $1/Z$ the relation \eqref{eq:xxxx2} is the Ward identity for a conserved charge with form-factor $\lambda_\rho (k)$. We recall that $\Lambda_\rho$ is defined without the factor $\lambda_\rho (k_F)$.

We plug these results into Eq.~\eqref{eq:cont-susc-par}, take the limit
$\Omega \gg v_F q \to 0$, and obtain,
\begin{equation}
  \label{eq:chi-j-determined}
  (\Lambda_JZ)^2\chi_{J,qp}(\q \to 0,0) =
  -
  \frac{\Omega^2}{q^2} \chi_{\rho,qp}(\frac{q}{\Omega} \to 0).
  \end{equation}
We showed in Sec. \ref{sec:2} that for any $l$, $\chi^{c(s)}_{l, qp}(\frac{q}{\Omega} \to 0)$ scales as $q^2/\Omega^2$, and the prefactor is expressed in terms of Landau parameters and is not singular.  Assuming that this holds for the conserved charge, we find that $(\Lambda_JZ)^2\chi_{J,qp}(\q \to 0,0)$  remains finite when Landau parameters change and pass through $-1$. Eq. \eqref{eq:chi-j-determined} then implies that there is no Pomeranchuk instability in the $J$ channel. It also explicitly connects $\Lambda_\rho,\Lambda_J, Z, m^*/m$ and $\chi_{\rho,qp},\chi_{J,qp}$ via Eqs. \eqref{eq:full-rho}+\eqref{eq:full-j}. This is the essence of our argument that the continuity equation implies constraints that connect low- and high- energy properties of the FL.

For the specific case of spin and charge density order parameters, one can easily verify that $\r^c (q)$ and $\r^s (q)$ commute with $H_{int}$ so the current density is bilinear in the creation and annihilation operators:
\begin{align}
  \label{eq:j-def-bilin}
  \J^{c}(\q,t) &= \frac{1}{m}\sum_{\k,\alpha}\k c^\dagger_{\k-\q/2,\alpha}c_{\k+\q/2,\alpha}, \\
  \J^{s}_m(\q,t) &= \frac{1}{m}\sum_{\k,\alpha\beta}\sigma^{\alpha\beta}_m\k c^\dagger_{\k-\q/2,\alpha}c_{\k+\q/2,\beta}.
\end{align}
In this case, the susceptibilities of $\r\cs,\J\cs$ correspond precisely to $\chi_{l=0}$ and $\chi_{l=1}$:
\begin{equation}
  \label{eq:rho-j-susc-l}
  \chi\cs_\rho = \chi_{l=0} \cs,\quad \chi\cs_J = \chi_{l=1} \cs.
\end{equation}
 Eq. \eqref{eq:cont-susc-par} then implies
\begin{equation}
  \label{eq:cont-susc-l01}
  (\Omega/q)^2\chi_{l=0}\cs(\tq) =
  \frac{1}{m^2} \hat q\cdot \left[\chi\cs_{l=1}(\tq) -\chi\cs_{l=1}(\q,0) \right] \cdot \hat q.
\end{equation}

Taking the $\Omega \gg v^*_F q$ limit, we obtain
\begin{equation}
  \label{eq:chi-0-coh-expansion}
  \chi_{l=0}\cs
  =
  -
  \chi_{l=0,0}\frac{v_F}{v_F^*}\left(\frac{v_F^*|\q|}{\Omega}\right)^2(1+F_1\cs) + O(|\q|^4/\Omega^4)
\end{equation}
Plugging the result into Eq. \eqref{eq:chi-j-determined} yields,
\begin{equation}
  \label{eq:charge-current-coh-equal}
  (\Lambda^{c,s}_{l=1}Z)^2\frac{v_F}{v_F^*}{}\frac{1}{1+F_1^{c,s}} = \frac{v_F^*}{v_F}(1+F_1\cs),
\end{equation}
i.e.,
\begin{equation}
  \label{eq:lambda-1-z-constraint}
  \Lambda_{l=1}\cs Z =\frac{v_F^*}{v_F}(1+F_1\cs),
\end{equation}
which is Eq. \eqref{n2}.

For the currents of conserved charge and spin there exists another constraint imposed by the longitudinal sum rule\cite{Ehrenreich1967,Leggett1965}:
\begin{equation}
  \label{eq:chi-j-sum-rule}
  \chi_J^\parallel(\q,0) = n/m
\end{equation}
where $n$ is the number density. The longitudinal sum rule is analogous to the longitudinal f-sum rule for the imaginary part of the inverse dielectric function \cite{Mahan} and can be derived from the gauge-invariance of the electromagnetic field \cite{Ehrenreich1967}. It is exact for a system where the electric current is proportional to the momentum density (with or without Galilean invariance), which is the case for any model of the form of Eqs. \eqref{n_1}, \eqref{n_5} with or without external potential $V(r)$. In effective low energy models (e.g. on a lattice), it is only  approximately correct\cite{Kiselev2017}. In either case, its implication is that the total $\chi_J$ is also finite.

\subsection{Conservation of momentum and $l=2$ susceptibility}
\label{sec:cons-moment-l=2}

Finally, we address the issue of the implication of the continuity equation for momentum in a Galilean invariant system. In this section we will refer to the momentum density by the symbol $\rho \equiv \rho_i$, and to the energy tensor by $J \equiv J_{ij}$ where $i,j$ denote spatial indices.

In Sec. \ref{sec:cont-equat-susc} we did not specify the nature of charge density and current. Thus, eq. \eqref{eq:cont-susc} is equally valid for the momentum densities and currents, the only change being that $\chi_\rho = \chi_\rho^{ij}(\q,\Omega)$ is a rank-2 symmetric tensor, and so is $(\chi_J^\parallel)^{ij} = (\hat q \cdot \chi_{J}(\q,\Omega)\cdot \hat q)^{ij}$. In the same manner, all arguments relating high frequency behavior of $\chi_\rho$ with the static behavior of $\chi_J$ go through, leading to Eq. \eqref{eq:chi-j-determined}. Thus
 $(\Lambda_J Z)^2\chi_{J,qp}$ is fully determined by $\chi_\rho$
and furthermore is always finite.

However, we now demonstrate that $J_{ij}$ cannot, in general, be expressed as a bilinear operator in $c^\dagger,c$. As a result, $\chi_J^\parallel$ does not have a simple relationship with $\chi_l$, e.g. with $\chi_{l=2}$. To see this, it is enough to examine the Hubbard model, i.e. take $U(|\q|) = U$ in Eq. \eqref{n_5}. The current operator Eq. \eqref{eq:j-def-bilin} has the following equation of motion,
\begin{align}
  \label{eq:dJ-dt}
  \frac{\partial\hat\rho(\q,t)}{\partial t} = -i\q \cdot\hat J
\end{align}
where
\begin{align}
  \label{eq:J-momentum-def}
  \hat J = \hat{J}_{kin} + \hat{J}_{int},
\end{align}
with
\begin{equation}
  \label{eq:j2-defs}
  \q \cdot \hat J_{kin} = \left[\rho,H_{free}\right],\quad \q \cdot \hat J_{int} = \left[\rho, H_{int}\right]
\end{equation}
which gives,
\begin{align}
  \label{eq:j-kin-def}
  \hat J^{ij}_{kin} &= \frac{1}{m^2}\sum_{\k}k_ik_jc^\dagger_{\k-\q/2}c_{\k+\q/2},\\
  \hat J^{ij}_{int} &= \delta_{ij}\frac{U}{m^2}\sum_{\k}n(\k)n(\q-\k),
\end{align}
where $n(\k) = \sum_{\p}c^\dagger_{\p-\k/2}c_{\p+\k/2}$.
If we had had $\hat J_{int} = 0$, then indeed Eq. \eqref{eq:chi-j-determined} could be used to constrain the $l=0,l=2$ channels, both of which appear in $\hat J_{kin}$. However, as it is, while Eq. \eqref{eq:chi-j-determined} does constraint $\chi_J$ to be finite, by itself it does not constrain any specific $l$ channels.

\section{Perturbative calculations for the Hubbard model: charge-current ad spin-current order parameters.}
\label{sec:4}

In this section we perform perturbative analysis of Eq. \eqref{eq:full_full} for $l=1$ and Eq.
\eqref{eq:cont-susc-l01}. We have three goals in our calculation: the first is to  show how one can derive the continuity equation diagrammatically, the second is to verify the relations between $\Lambda^{c(s)}_{l=1} Z, \chi^{c(s)}_{l=1,inc}$ and $F^{c(s)}_{l=1}$, Eqs. \eqref{n2} and \eqref{n3}, in direct expansion in the interaction,
and
the third goal is to clarify the origin of the relation between high- and low- energy contributions to Eqs. \eqref{n2} and \eqref{n3}.

We proceed in three steps. First, we derive Eq. \eqref{eq:cont-susc-l01} diagrammatically to first-order in $U(\q)$. We will see that although there are no dynamical corrections to this order (i.e. $Z,\Lambda^{c(s)}_l = 1$), nevertheless self-energy corrections are crucial, indicating one should go beyond RPA. Then, we perform a combined analytical and numerical analysis of $\chi^{c(s)}_{l=1}$ at order $U^2$ for the Hubbard model, and explicitly verify Eqs. \eqref{n2}, \eqref{n3}. Going to to second order in $U$ is essential, because only at this order do contributions away from the FS begin to accumulate, see Sec. \ref{sec:higher-orders-uq}. Finally, we demonstrate  that the high-energy contributions to $\chi\cs_{l=1}$ can be re-expressed as low-energy ones, due to
a
special property of the sum of particle-hole and particle-particle bubbles.

\subsection{Diagrammatic derivation of the continuity equation}
\label{sec:first-order-uq}

In this subsection we show how Eq. (\ref{eq:cont-susc-l01}) can be reproduced in a diagrammatic calculation. Already at this order we will see that one needs to treat self-energy and vertex corrections on equal footings because the continuity equation emerges due to a particular cancellations between these two types of corrections.

To begin with, we re-write  Eq. \eqref{n_3} for free-fermion susceptibility for a current order parameter with $\lambda^{c(s)}_{l=1} (k) = {\bf k}\cdot \hat q$ as
\begin{align}
  \label{eq:chi1-zero-order}
  q^2\chi_{l=1,0}\cs(q)
  =-2\int\frac{d^2k}{(2\pi)^2}\frac{n_F(\xi_{\k-\frac{\q}{2}})-n_F(\xi_{\k+\frac{\q}{2}})}{\Omega-\frac{1}{m}\k\cdot\q+i\delta_{\Omega}}(\k\cdot\q)^2	\end{align}
Here and later on we omit the $\parallel$ symbol for clarity. We then rewrite the form factor as:	
\begin{equation}
  \label{eq:add-sub}
  (\k\cdot\q)^2=(\k\cdot\q+m\Omega)(\k\cdot\q-m\Omega)+m^2\Omega^2
\end{equation}
and obtain
\begin{align}
  q^2\chi\cs_{l=1,0}(q) &=  2m^2\int\frac{d^2k}{(2\pi)^2}[n_F(\xi_{\k-\frac{\q}{2}})-n_F(\xi_{\k+\frac{\q}{2}})] \nn\\
 &\qquad \times \left[-\Omega -\frac{1}{m}\k\cdot \q + \frac{ \Omega^2 }{\Omega-\frac{1}{m}\k\cdot\q+i\delta_\Omega}\right]
 \label{ll2}
\end{align}
The $\Omega$ term vanishes after integration over $k$.  The other two terms are easily identified as $q^2\chi_{l=1,0}\cs(\q,0)$ and  $\Omega^2\chi_{l=0,0}\cs(q)$, so that:
\begin{equation}
  \label{eq:chi-zero-order-cont}
  \frac{q^2}{m^2}\left(\chi\cs_{l=1,0}(q) - \chi\cs_{l=1,0}(\q,0)\right) = \Omega^2\chi\cs_{l=0,0}(q).
\end{equation}

We now use the same tactics for first order corrections to $\chi\cs_{l=1}$. The corresponding diagrams are given in Fig. \ref{fig:chi-1st-order}. Diagram \ref{fig:chi-1st-order}c, the RPA correction, gives
\begin{flalign}
  \label{eq:chi-bubble-u1}
  \frac{q^2}{m^2}\chi_{l=1,2c}\cs = \left[2\int\frac{d^2k}{(2\pi)^2}\frac{n_F(\xi_{\k-\frac{\q}{2}})-n_F(\xi_{\k+\frac{\q}{2}})}{\Omega-\frac{1}{m}\k\cdot\q+i\delta_{\Omega}}\k\cdot\q\right]^2\times\nn\\
  U(\q)
\end{flalign}+
By making use of
\begin{equation}\label{eq:n64}
	\frac{\k\cdot\q}{\Omega-\frac{1}{m}\k\cdot\q+i\delta_{\Omega}}=\frac{m\Omega}{\Omega-\frac{1}{m}\k\cdot\q+i\delta_{\Omega}}-m
\end{equation}
we find
\begin{equation}
  \label{eq:delta-chi-0-cont}
  \frac{q^2}{m^2}
  \chi_{l=1,2c}\cs(\tq) = \Omega^2 \chi_{l=0,2c}\cs(\tq)
\end{equation}
Note that there is no need to subtract the static part because $\chi_{l=1,2c}\cs({\bf q}, 0)$ vanishes.

For the remaining two diagrams in Fig. \ref{fig:chi-1st-order} we obtain
\begin{widetext}
  \begin{align} \label{eq:RPA-MT-corrections}
      q^2 \chi\cs_{l=1,2a}(\tq) &= -2\int\frac{d^2k}{(2\pi)^2}\frac{d^2p}{(2\pi)^2}\frac{[n_F(\xi_{\p-\frac{\q}{2}})-n_F(\xi_{\p+\frac{\q}{2}})] [n_F(\xi_{\k-\frac{\q}{2}})-n_F(\xi_{\k+\frac{\q}{2}})]}{(\Omega-\frac{1}{m}\k\cdot\q+i\delta_{\Omega})^2}U(|\p-\k|) (\k \cdot \q)^2\\
\label{eq:2b-finite-q}
    q^2 \chi\cs_{l=1,2b}(\tq) &= 2\int\frac{d^2k}{(2\pi)^2}\frac{d^2p}{(2\pi)^2}\frac{[n_F(\xi_{\p-\frac{\q}{2}})-n_F(\xi_{\p+\frac{\q}{2}})][n_F(\xi_{\k-\frac{\q}{2}})-
n_F(\xi_{\k+\frac{\q}{2}})]}{(\Omega-\frac{1}{m}\p\cdot\q+i\delta_{\Omega})(\Omega-\frac{1}{m}\k\cdot\q+i\delta_{\Omega})}U(\p-\k) (\k \cdot \q) (\p \cdot \q)
  \end{align}
\end{widetext}
Applying again \eqref{eq:n64} we find,
\begin{equation}
   \frac{q^2}{m^2}(\chi_{l=1,2a}\cs(\tq)+\chi_{l=1,2b}\cs(\tq)) = \Omega^2 (\chi_{l=0,2a}\cs(\tq)+\chi_{l=0,2b}\cs(\tq))
   \label{ll1}
 \end{equation}
 The static part of the sum of the two contributions cancel out. Eqs. \eqref{eq:delta-chi-0-cont} and \eqref{ll1} verify Eq. \eqref{eq:cont-susc-l01} to order $U$.

 We emphasize that $\chi\cs_{l=1,2a}(\tq)$ and $\chi\cs_{l=1,2b}(\tq)$,  when taken separately, do not satisfy the continuity equation \eqref{eq:cont-susc-l01}, and only the sum of the two terms obeys (\ref{ll1}).  This is an indication that, within diagrammatics, the continuity equation emerges due to fine cancellations between self-energy and vertex corrections, and one
 should go beyond RPA at each order in $U$ to reproduce it.

\begin{figure*}
  \centering
  \includegraphics[width=17cm]{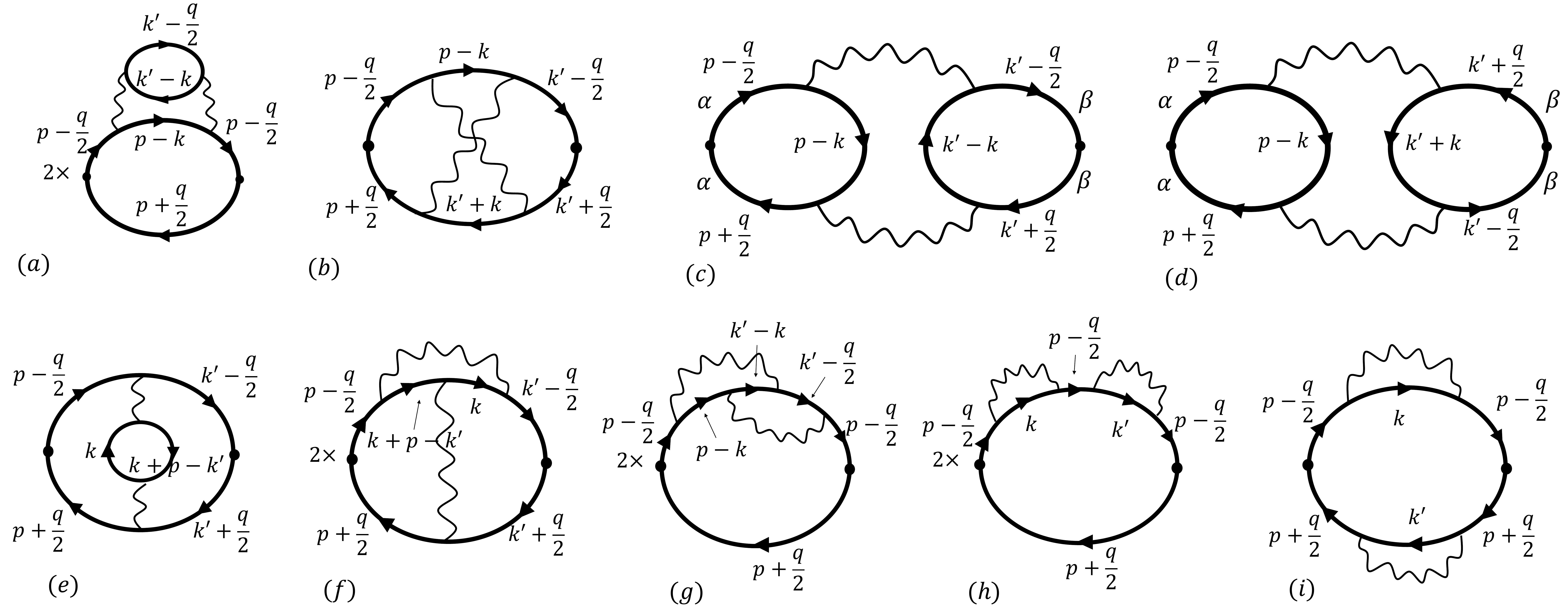}
  \caption{Diagrams in second order of $U$. For constant interaction, (e) and (f), (h) and (i) cancels out. (g) half cancels (a) and (b) half cancels (d). What remains are half of (a), (c),  and half of (d).}\label{fig:2nd-order}
\end{figure*}

\subsection{Evaluation of $\chi_{l=1}\cs$ to order $U^2$}
\label{sec:eval-chi_l-1cs-order}

We now present the results of explicit calculations of the static susceptibilities to order
$U^2$. We identify contributions to $\chi^{c(s)}_{l=1,inc}$ and $\left(\Lambda^{c(s)}_{l=1} Z\right)^2$, and $\chi^{c(s)}_{l,qp}$ from each diagram, and compute them
 by a combination of analytical and numerical methods.
We also independently compute the vertex renormalization $\Lambda\cs_{l=1}$ to  order $U^2$.

There are nine different diagrams for the current susceptibility to second order in $U(q)$, see Fig. \ref{fig:2nd-order}.  To simplify the numerics, we approximate $U(q)$ by a constant $U$  i.e., consider $U^2$ renormalizations in the Hubbard model. For a constant $U$, Landau parameters $F^{c(s)}_{l}$ also only emerge at order $U^2$, i.e., the incoherent part of the susceptibility, vertex renormalizion, renormalization of the quasiparticle $Z$, and Landau parameters are all of order $U^2$.  We make use of previously known results \cite{Chubukov}
\begin{equation}
  \label{eq:mmstar-z}
  F_1^c=-F_1^s=\frac{m^2U^2}{8\pi^2},\qquad Z = 1 - 1.39 \frac{m^2U^2}{8\pi^2},
 \end{equation}
 and
\begin{equation}
  \label{eq:mstar-m}
  \frac{m^*}{m} = 1 + F_1^c = 1 + \frac{m^2U^2}{8\pi^2}
\end{equation}
which  holds for a Galilean-invariant system~\cite{AGD,Lifshitz1980}.
The order $U^2$ is the first one in perturbative expansion at which differences between the charge and spin channels emerge, in the form of the Aslamazov-Larkin (AL) diagrams, Figs. \ref{fig:2nd-order}c,d. The AL diagrams contribute in the charge channel and vanish in the spin channel, as can be seen from direct spin summation.

Consider the charge channel first.  It is straightforward to  identify the diagrams  in Figs. \ref{fig:2nd-order}, which give equal contributions, up to overall factor.
One can easily verify that
$\chi_{6a} = -2 \chi_{6g},~ \chi_{6d} = -2 \chi_{6b}$,  and $\chi_{6e} = -\chi_{6f}$.
In addition, using the relation
\begin{equation}\label{eq:g-relation}
  \int d\omega_{p} (G_{\tp-\frac{\tq}{2}}^0)^3G^0_{\tp+\frac{\tq}{2}}=-\frac{1}{2}\int d\omega_p (G^0_{\tp-\frac{\tq}{2}}G^0_{\tp+\frac{\tq}{2}})^2,
\end{equation}
we  find $\chi_{6h} = -\chi_{6i}$.
In Eq. \eqref{eq:g-relation} and throughout this section we
 denote $G_0(k) \equiv G^0_k$ for compactness.
Summing up the contributions to the charge-current susceptibility, we obtain at order $U^2$,
\beq
\delta\chi^c_{l=1} = \frac{1}{2} \chi_{6a} + \chi_{6c} + \frac{1}{2} \chi_{6d}
\label{ll3}
\eeq
A similar consideration for the spin susceptibility yields
\beq
\delta\chi^s_{l=1} = \frac{1}{2} \left(\chi_{6a} - \chi_{6d}\right)
\label{ll4}
\eeq
In explicit form
\begin{widetext}
\begin{equation}\label{eq:all-diagrams}
  \begin{aligned}
  \chi_{6a}&=
  8 U^2 \int \frac{d^3k d^3k' d^3p}{(2\pi)^9} (\p\cdot\hat{q})^2                  (G^0_{\tp-\frac{\tq}{2}})^2G^0_{\tp+\frac{\tq}{2}}G^0_{\tp-\tk}G^0_{\tk'-\tk}G^0_{\tk'-\frac{\tq}{2}},\\
  \chi_{6c}&=
  4 U^2 \int \frac{d^3k d^3k' d^3p}{(2\pi)^9}   (\p\cdot\hat{q}) (\k'\cdot\hat{q})
  G^0_{\tp-\frac{\tq}{2}}G^0_{\tp+\frac{\tq}{2}}G^0_{\tk'-\frac{\tq}{2}}G^0_{\tk'+\frac{\tq}{2}}G^0_{\tp-\tk}G^0_{\tk'-\tk},\\
  \chi_{6d}&=
  4 U^2 \int \frac{d^3k d^3k' d^3p}{(2\pi)^9}  (\p\cdot\hat{q}) (\k'\cdot\hat{q})                   G^0_{\tp-\frac{\tq}{2}}G^0_{\tp+\frac{\tq}{2}}G^0_{\tk'-\frac{\tq}{2}}G^0_{\tk'+\frac{\tq}{2}}G^0_{\tp-\tk}G^0_{\tk'+\tk}.
  \end{aligned}
\end{equation}
\end{widetext}
We set $\Omega=0$ and take ${\bf q}$ to be small but finite. After integration over frequency, we split each diagram into three parts: ``high'', ``middle'', and ``low'' (which we label ``H'', ``M'', and ``L''), depending on whether zero, one, or two internal fermionic momenta  are confined to the FS, e.g. $\chi_{6a}=\chi^{H}_{6a}+\chi^M_{6a}+\chi^L_{6a}$.
 In this computational scheme, AL diagrams contain ``H'', ``M'', and ``L'' parts, while the diagram with self-energy renormalization contains ``H'' and ``M'' parts.  In explicit form we have
\begin{widetext}
\begin{subequations}\label{eq:all}
  \begin{align}
                       \chi^H_{6a}&= -8 U^2 \int \frac{d^2k d^2k' d^2p}{(2\pi)^6} (\p\cdot\hat{q})^2
                                       \frac{[n_F(\xi_{\bm{k'}})-n_F(\xi_{\bm{k'-k}})] [n_F(\xi _{\bm{p}})-n_F(\xi _{\bm{p-k}})][n_B(\xi _{\bm{k'}}-\xi _{\bm{k'-k}})-n_B(\xi _{\bm{p}}-\xi _{\bm{p-k}})]}{(\xi _{\bm{k'}}-\xi _{\bm{k'-k}}-\xi _{\bm{p}}+\xi _{\bm{p-k}})^3}\\
    \chi^H_{6c}&= +8 U^2 \int \frac{d^2k d^2k' d^2p}{(2\pi)^6} (\p\cdot\hat{q}) (\k'\cdot\hat{q})
                    \frac{[n_F(\xi_{\bm{k'}})-n_F(\xi_{\bm{k'-k}})] [n_F(\xi _{\bm{p}})-n_F(\xi _{\bm{p-k}})][n_B(\xi _{\bm{k'}}-\xi _{\bm{k'-k}})-n_B(\xi _{\bm{p}}-\xi _{\bm{p-k}})]}{(\xi _{\bm{k'}}-\xi _{\bm{k'-k}}-\xi _{\bm{p}}+\xi _{\bm{p-k}})^3}\\
    \chi^H_{6d}&= -8 U^2 \int \frac{d^2k d^2k' d^2p}{(2\pi)^6} (\p\cdot\hat{q}) (\k'\cdot\hat{q})                     
                    \frac{[n_F(\xi_{\bm{k'}})-n_F(\xi_{\bm{k'+k}})] [n_F(\xi _{\bm{p}})-n_F(\xi _{\bm{p-k}})][n_B(-\xi _{\bm{k'}}+\xi _{\bm{k'+k}})-n_B(\xi _{\bm{p}}-\xi _{\bm{p-k}})]}{(\xi _{\bm{k'}}-\xi _{\bm{k'+k}}+\xi _{\bm{p}}-\xi _{\bm{p-k}})^3} \\
    \chi^M_{6a}&= -4 U^2 \int \frac{d^2k d^2k' d^2p}{(2\pi)^6}  (\p\cdot\hat{q})^2
                     \left(1+\frac{|\bm{k'}|\cos\phi_{\bm{k'}}}{|\bm{p}|\cos\phi_{\bm{p}}}\right)n'_F(\xi_{\bm{p}})\frac{[n_F(\xi_{\bm{k'}})-n_F(\xi_{\bm{k'-k}})] [n_B(\xi _{\bm{k'}}-\xi _{\bm{k'-k}})-n_B(\xi _{\bm{p}}-\xi _{\bm{p-k}})]}{(\xi _{\bm{k'}}-\xi _{\bm{k'-k}}-\xi _{\bm{p}}+\xi _{\bm{p-k}})^2}\\
    \chi^M_{6c}&= +8 U^2 \int \frac{d^2k d^2k' d^2p}{(2\pi)^6} (\p\cdot\hat{q}) (\k'\cdot\hat{q})
                     n'_F(\xi_{\bm{p}})\frac{[n_F(\xi_{\bm{k'}})-n_F(\xi_{\bm{k'-k}})] [n_B(\xi _{\bm{k'}}-\xi _{\bm{k'-k}})-n_B(\xi _{\bm{p}}-\xi _{\bm{p-k}})]}{(\xi _{\bm{k'}}-\xi _{\bm{k'-k}}-\xi _{\bm{p}}+\xi _{\bm{p-k}})^2}\\
    \chi^M_{6d}&= +8 U^2 \int \frac{d^2k d^2k' d^2p}{(2\pi)^6}  (\p\cdot\hat{q}) (\k'\cdot\hat{q})
                     n'_F(\xi_{\bm{p}})\frac{[n_F(\xi_{\bm{k'}})-n_F(\xi_{\bm{k'+k}})] [n_B(-\xi _{\bm{k'}}+\xi _{\bm{k'+k}})-n_B(\xi _{\bm{p}}-\xi _{\bm{p-k}})]}{(\xi _{\bm{k'}}-\xi _{\bm{k'+k}}+\xi _{\bm{p}}-\xi _{\bm{p-k}})^2}\\
    \chi^L_{6a}&= +4 U^2 \int \frac{d^2k d^2k' d^2p}{(2\pi)^6} (\p\cdot\hat{q})^2
                     \frac{|\bm{k'}|\cos\phi_{\bm{k'}}}{|\bm{p}|\cos\phi_{\bm{p}}}n'_F(\xi_{\bm{p}})n'_F(\xi_{\bm{k'}})\frac{n_B(\xi _{\bm{k'}}-\xi _{\bm{k'-k}})-n_B(\xi _{\bm{p}}-\xi _{\bm{p-k}})}{\xi _{\bm{k'}}-\xi _{\bm{k'-k}}-\xi _{\bm{p}}+\xi _{\bm{p-k}}}\\
    \chi^L_{6c}&= -4 U^2 \int \frac{d^2k d^2k' d^2p}{(2\pi)^6} (\p\cdot\hat{q}) (\k'\cdot\hat{q})
    n'_F(\xi_{\bm{p}})n'_F(\xi_{\bm{k'}})\frac{n_B(\xi _{\bm{k'}}-\xi _{\bm{k'-k}})-n_B(\xi _{\bm{p}}-\xi _{\bm{p-k}})}{\xi _{\bm{k'}}-\xi _{\bm{k'-k}}-\xi _{\bm{p}}+\xi _{\bm{p-k}}}\\
    \chi^L_{2d}&= -4 U^2 \int \frac{d^2k d^2k' d^2p}{(2\pi)^6}  (\p\cdot\hat{q}) (\k'\cdot\hat{q})
    n'_F(\xi_{\bm{p}})n'_F(\xi_{\bm{k'}})\frac{n_B(-\xi _{\bm{k'}}+\xi _{\bm{k'+k}})-n_B(\xi _{\bm{p}}-\xi _{\bm{p-k}})}{\xi _{\bm{k'}}-\xi _{\bm{k'+k}}+\xi _{\bm{p}}-\xi _{\bm{p-k}}}
  \end{align}
\end{subequations}\end{widetext}
Here $n_B(\xi) = -\Theta(-\xi)$ at $T=0$.

The ``H'' contributions can be
 evaluated by just setting $\Omega=0$ and $\q=0$ in Eq.\eqref{eq:all-diagrams}, e.g.
\begin{equation}
  \chi^H_{6a}=8 U^2 \int \frac{d^3k d^3k' d^3p}{(2\pi)^9} {\bm{p}}^2
  (G^0_{\tp})^3G^0_{\tp-\tk}G^0_{\tk'-\tk}G^0_{\tk'}
\end{equation}
The sum of ``H'' parts is then the incoherent part  of the susceptibility
 \begin{equation}
  \delta \chi_{l=1}^{c(s),H} = \chi_{l=1,inc}\cs
\end{equation}
 The  ``M'' and ``L''  parts  determine
 \begin{eqnarray}\label{lambdaz}
   && \delta \chi_{l=1}^{c(s),M} = \chi_{l=1,0}\cs  \left( \frac{m^*}{m} \left(\Lambda^{c(s)}_{l=1} Z\right)^2 -1\right)  \nonumber \\
   &&  \delta \chi_{l=1}^{c(s),L} =  - \chi^{c(s)}_{l=1,0} F^{c(s)}_{l=1}
  \end{eqnarray}
The ``L'' part can be computed analytically and yields
\begin{equation}
  \label{eq:bl-al1l-res}
  \chi^L_{6a} = \chi^L_{6c} = 0,
  \end{equation}
and
\begin{equation}
  ~\chi^L_{6d} =  -\frac{m^2U^2}{4\pi^2}\chi_{l=1,0}
\end{equation}
where $\chi_{l,0}$ is a free-fermion susceptibility, given by \eqref{n_14}.
Using (\ref{ll3}), (\ref{ll4}) and (\ref{eq:mmstar-z}), we find $\delta \chi_{l=1}^{c(s),L} =  - \chi^{c(s)}_{l=1,0} F^{c(s)}_{l=1}$ as in (\ref{lambdaz}).
The ``M'' and ``H'' terms in Eq.~\eqref{eq:all} are high dimensional principal value integrals, which we evaluate numerically. Details of our numerics can be found in the Appendix.

According to Eqs.  \eqref{n22} and \eqref{n3},  the
total ``H'' contributions to  charge-current susceptibility, $\delta \chi^c_{l=1,H}$  should vanish, while other contributions should obey, to order $U^2$,
\begin{eqnarray}
  \label{eq:deviations-cs}
  &&\delta \chi^c_{l=1,M} =  \left(\frac{m}{m^*} (1+ F^c_{1})^2 -1\right) \chi_{l=1,0} \approx F^c_{1} \chi_{l=1,0} \nonumber \\
  &&\delta \chi^s_{l=1,M} = \left(\frac{m}{m^*} (1+ F^s_1)^2 -1\right) \chi_{l=1,0} \approx (2 F^s_{1} - F^c_1) \chi_{l=1,0} \nonumber \\
  &&\delta \chi^s_{l=1,H} =  \left(\frac{m^*}{m} -1 - F^s_{l}\right) \chi_{l=1,0} \approx (F^c_{1} - F^s_1) \chi_{l=1,0}
\end{eqnarray}
Using Eq. (\ref{eq:mmstar-z}) for $F^{c(s)}_1$  and $\chi_{l=1,0} = mk^2_F/(2\pi)$  (recall that for spin and charge currents $f^{c(s)}_{l=1} (k_F) =1$), we obtain
\def\chib{\overline{\chi}}
\begin{equation}\label{f1}
   \begin{aligned}
     &\delta \chi^c_{l=1,M} = \chib,\qquad  \delta \chi^c_{l=1,H} =0, \\
     &\delta\chi^s_{l=1,M} = -3\chib,  \quad\delta \chi^s_{l=1,H} =2\chib,
   \end{aligned}
 \end{equation}
 where $\chib = m^3U^2 k^2_F/16\pi^3$.
 In Table  \ref{tab:delta-chi} we list $\delta \chi^{c(s)}_{l=1,H}$, $\delta \chi^{c(s)}_{l=1,M}$ and $\delta\chi_{l=1,L}\cs$ in units of $\chib$.
 \begin{figure}
	\includegraphics[width=8cm]{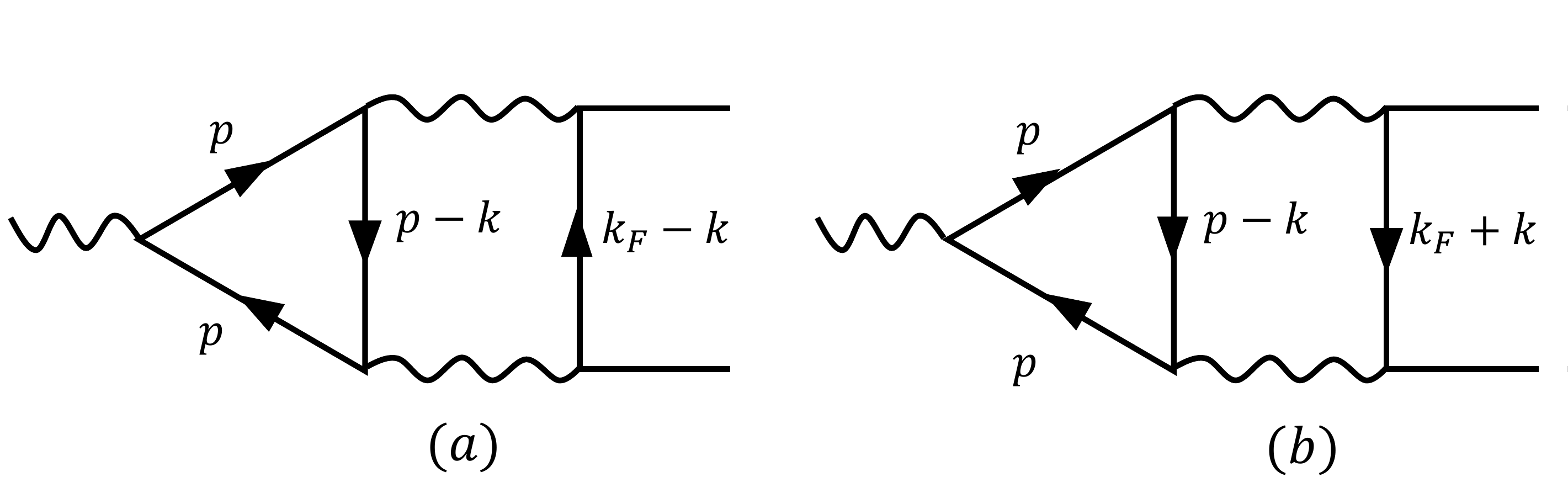}
	\caption{
The two AL vertex correction diagrams for  three-leg vertex.
}\label{fig:3vertex}
\end{figure}
\begin{figure}
  \includegraphics[width=8cm]{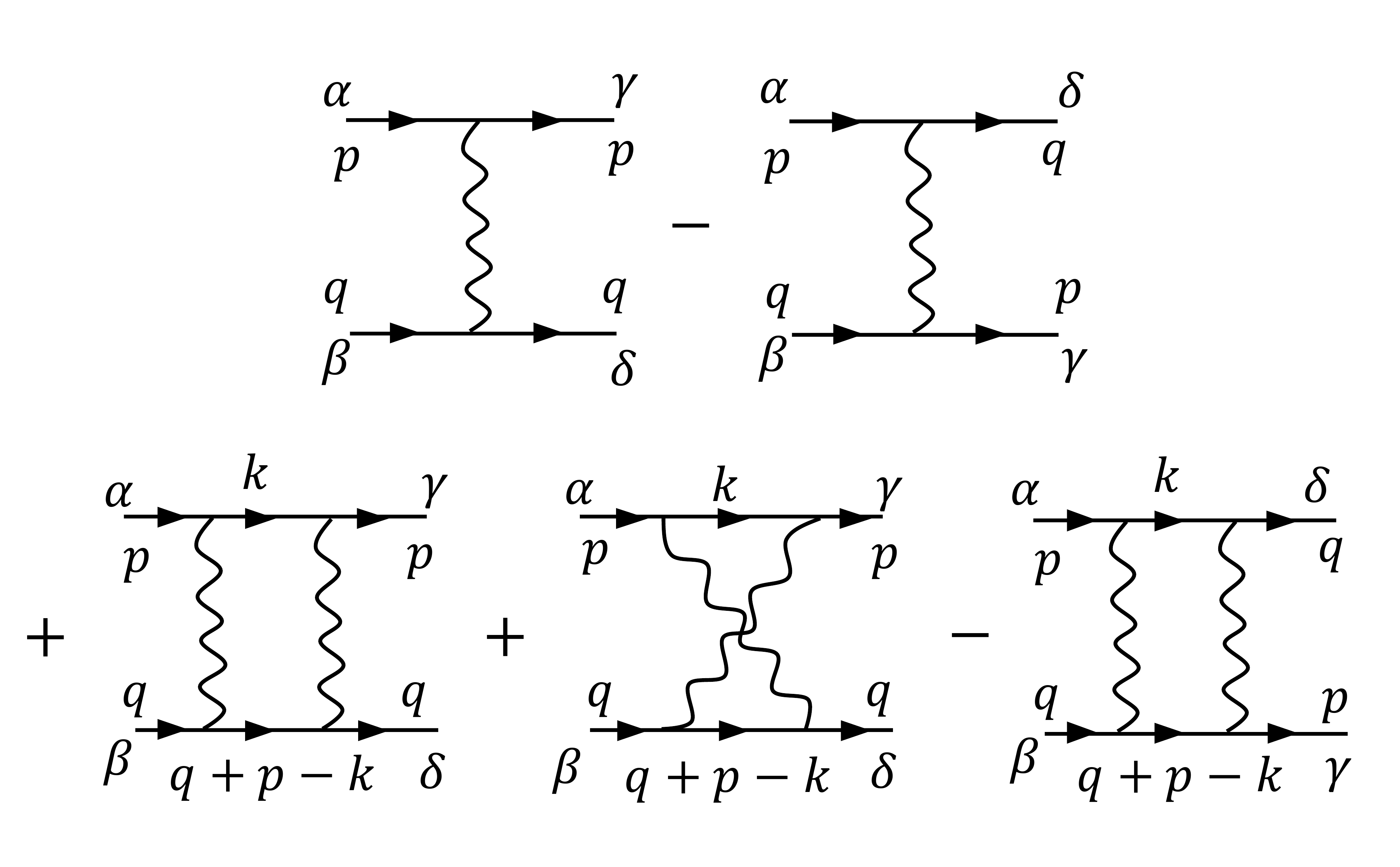}
  \caption{Diagrams for $\Gamma^{\omega}$ to second order, for a constant $U$.
  }\label{fig:4vertex}
\end{figure}
We also computed $\Lambda^{c(s)}_{l=1}$ independently, by collecting vertex correction diagrams, keeping external particles at the FS.  Applying the same tactics as before, i.e., identifying equivalent contributions to reduce the number of diagrams, we find that
\bea
&&\Lambda^{c}_{l=1}  = 1+ \Lambda_{7a} + \frac{1}{2}\Lambda_{7b} \nonumber \\
&& \Lambda^{s}_{l=1}  =  1 - \frac{1}{2}\Lambda_{7b}
\eea
where $\Lambda_{7a}$ and $\Lambda_{7b}$ are two vertex corrections in Fig.\ref{fig:3vertex}.
In explicit form
\begin{equation}\label{eq:3vertex}
  \begin{aligned}
  \Lambda_{7a}&=-\frac{2U^2}{k_F} \int \frac{d^3k d^3p}{(2\pi)^6}(\p\cdot\hat{q})(G^0_{\tp})^2G^0_{\tp-\tk}G^0_{k_F\hat n-\tk}\\
\Lambda_{7b}&=-\frac{2U^2}{k_F}\int \frac{d^3k d^3p}{(2\pi)^6}(\p\cdot\hat{q})(G^0_{\tp})^2G^0_{\tp-\tk}G^0_{k_F\hat n+\tk}
  \end{aligned}
\end{equation}
where $ \hat n$ is a unit vector. We evaluated the integrals in the RHS of  \eqref{eq:3vertex} numerically and the results are presented in Table \ref{tab:delta-lambda}.
From (\ref{eq:deviations-cs}) and  (\ref{eq:mmstar-z}) we expect
\begin{equation}
  \label{eq:delta-gamma-z-expected}
  \Lambda^c_{l=1} - 1 =1.39 \overline{\Lambda},\quad   \Lambda^s_{l=1} - 1 = -0.61\overline{\Lambda}
\end{equation}
where $\overline{\Lambda} = \frac{m^2U^2}{8\pi^2}$. We see that these relations are satisfied, as they should be.
\begin{table*}
  \begin{minipage}[t]{0.45\hsize}
  \centering
  \begin{tabular}{lcccc}
    \hline\hline
    channel\quad& $\delta\chi_{l=1,L}$ &$\delta\chi_{l=1,M}$ & $\delta\chi_{l=1,H}$ & $\delta\chi_{l=1}$\\
    \hline
    charge & $-1$ & $+0.99 \pm 0.02$ &$+0.01 \pm 0.04$&  $\mathbf{0.01 \pm 0.04}$ \\
    spin & $+1$  & $-2.97 \pm 0.02$ & $+1.98 \pm 0.02$&  $\mathbf{0.01 \pm 0.02}$ \\
    \hline\hline
  \end{tabular}
  \caption{
    The contributions to susceptibilities for charge-current and spin-current order parameters ($l=1$ orders with form factor $\lambda\cs_{l=1} (p)={\bf p}$) at order $U^2$, from fermions at high (``H''),  middle (``M'') and low (``L'') energies. The ``L'' contribution
    was obtained analytically,
    and the
    ``M'' and ``H'' contributions
    were
    obtained numerically.
    The numbers are in units of $\chib = m^3U^2 k^2_F/16\pi^3$.
    The results agree with Eq. \eqref{f1} and, hence, with Eqs.  \eqref{n22} and \eqref{n3}.
   \label{tab:delta-chi}}
\end{minipage}
\hfill
\begin{minipage}[t]{0.45\hsize}
  \centering

  \begin{tabular}{lcc}
    \hline\hline
    channel & from Eq.\eqref{eq:3vertex} & from $\delta\chi_{l=1,M}$\\
    \hline
    charge & $ +1.39 \pm 0.02$ & $+1.38 \pm 0.01$\\
    spin & $-0.604 \pm 0.008 $ & $-0.60 \pm 0.01$\\
    \hline\hline
  \end{tabular}
  \caption{Numerical results for $\Lambda_{l=1}\cs-1$ for the case when the form factor is $\lambda\cs_{l=1} (p)= {\bf p}$. The results are in units of $\overline{\Lambda} = \frac{m^2U^2}{8\pi^2}$. The first column is obtained from a direct evaluation of Eq.\eqref{eq:3vertex} and the second one is extracted from our calculation of $\delta\chi\cs_{l=1,M}$, via Eq.\eqref{lambdaz} . The results are in agreement with Eq.\eqref{eq:delta-gamma-z-expected} and, hence, with Eq.\eqref{n2}.}
  \label{tab:delta-lambda}
  \end{minipage}
\end{table*}

\subsection{Microscopic explanation for the absence of $l=1$ Pomerachuk instabilities}
\label{sec:micr-expl-absence}

We now present microscopic arguments as to why $\Lambda Z$ and $\chi_{l=1, inc}$, for charge-current and spin-current susceptibilities are expressed via Landau parameters. We will analyze Eq. \eqref{eq:lambda-1-z-constraint} for the spin channel, where $\Lambda Z = 1 + F^s_1-F^c_1$.

The quasiparticle residue $Z$ can be expressed via $\Gamma^\omega_{\alpha \beta, \alpha \beta}$ using a Ward identity for any conserved "charge" \cite{AGD,Lifshitz1980,Peskin1995}. For our purpose it is best to use the Ward identity associated with conservation of total momentum (recall that we consider a Galilean invariant system).  Substituting ${\lambda_\rho (k} = {\bf k}$ into (\ref{eq:xxxx2}) we obtain
\begin{equation}
  \label{eq:Z-GI-FL}
  \frac{1}{Z} = 1
  -  \frac{i}{2 k_F}\sum_{\alpha\beta}\int \frac{d^3q}{(2\pi)^3} \Gamma^\omega_{\alpha\beta,\alpha\beta}(k_F\hat p, \q)(G_q^2)^\omega \hat p \cdot \q
\end{equation}
The  renormalization of the spin-current  vertex can be written as
\begin{align}
  \label{eq:delta-lambda-1s-FL}
  &\Lambda_{l=1}^s \sigma^z_{\beta\beta} = \sigma^z_{\beta\beta} - \nn \\
  &\qquad\qquad\frac{i}{k_F}\sum_{\alpha}
  \int \frac{d^3q}{(2\pi)^3}
  \Gamma^\omega_{\alpha\beta,\alpha\beta}(k_F\hat p, \q)(G_q^2)^\omega \hat p \cdot \q \sigma^z_{\alpha\alpha}
\end{align}
where $\hat p \cdot \q$ is now simply the form-factor for the current. The vertex function $\Gamma^\omega$ to order $U^2$ is given by the diagrams in Fig. \ref{fig:4vertex}.
In explicit form
\begin{widetext}
\begin{align}
  \label{eq:gamma-w-u2}
  \Gamma^\omega_{\alpha\beta,\gamma\delta}(p=(k_F\hat p, 0),q) &= \frac{1}{2}\delta_{\alpha\gamma}\delta_{\beta\delta}\left[U + i U^2 \int \frac{d^3k}{(2\pi)^3}\left(2 G_k G_{q-p + k} + G_k G_{q+p-k}\right)\right] \nn \\
  &\qquad\qquad -\frac{1}{2}\sigma_{\alpha\gamma}\cdot\sigma_{\beta\delta}\left[U + i U^2 \int \frac{d^3k}{(2\pi)^3} G_k G_{q+p-k}\right].
\end{align}
\end{widetext}

Summing up contributions from both $Z$ and $\Gamma$ we obtain, to order $U^2$,
\begin{align}
  \label{eq:delta-lambdaZ-u2}
  &\Lambda^s_{l=1} Z =  1 - \nn \\
   &\qquad\quad\frac{U^2}{k_F} \int \frac{d^3kd^3q}{(2\pi)^6}\left(G_k G_{q-p + k} + G_k G_{q+p-k}\right)\hat p \cdot \q (G_q^2)^\omega
\end{align}
As written, the integral in the RHS of  Eq. \eqref{eq:delta-lambdaZ-u2} is not confined to the FS. However, the sum can be re-expressed as an integral over the FS. The reason for this is the identity \cite{Chubukov},
\begin{flalign}
  \label{eq:ph-pp-ident}
  \int \frac{d^3kd^3q}{(2\pi)^6}\left(G_k G_{q-p + k} + G_k G_{q+p-k}\right)(G_{q+p_\epsilon} - G_q)=0,
\end{flalign}
where $p_\epsilon = \epsilon(k_F\hat p,\Omega)$ and $\epsilon \to 0$. This identity can be proven  by a simple relabeling of indices on the p-h bubble. Choosing $\Omega = 0$ and expanding to first order in $\epsilon$, we obtain
\begin{equation}
  \int \frac{d^3kd^3q}{(2\pi)^6}\left(G_k G_{q-p + k} + G_k G_{q+p-k}\right)(G_q^2)^k\hat p \cdot \q=0,
 \label{eq:ph-pp-connect}
\end{equation}
where $(G_q^2)^k= \mbox{lim}_{\k\to 0}G(\q+\k,\Omega)G(\q,\Omega)$. This $(G_q^2)^k$ has a regular piece, equal to $(G_q^2)^\omega$, and an extra piece which comes from the FS. Using the known relation~\cite{AGD}
\begin{equation}
(G_q^2)^k = (G_q^2)^\omega - \frac{2 \pi i Z^2}{v_F^*}
 \delta(\omega)\delta(|\q|-k_F),
\label{ll7}
\end{equation}
substituting into (\ref{eq:ph-pp-connect}), and using Eq. (\ref{eq:gamma-w-u2}) to extract the Landau parameters, we obtain
\begin{widetext}
\beq
\label{llla}
 \frac{U^2}{k_F}\int \frac{d^3kd^3q}{(2\pi)^6}\left(G_k G_{q-p + k} + G_k G_{q+p-k}\right)(G_q^2)^\omega\hat p \cdot \q =
 \int \frac{d\theta}{2\pi} \left( F^c (\theta) - F^s (\theta) \right) \cos{\theta}  = F^c_{l=1} - F^s_{l=1}
 \eeq
\end{widetext}
Substituting into (\ref{eq:delta-lambdaZ-u2}), we recover  Eq. \eqref{eq:delta-gamma-z-expected}.

We emphasize that only the product $ \Lambda^s_{l=1} Z$ is expressed via the integral over the FS.  Taken separately, $ \Lambda^s_{l=1}$ and $Z$ are determined by integrals which are not confined to the FS. We also note that the same Eq. (\ref{llla}) allows one to express the effective mass, computed to order $U^2$ in a direct perturbation theory,  as the integral over the FS in Eq. (\ref{eq:mstar-m}) (see Ref. ~\cite{Chubukov} for details).

\section{
Arbitrary form-factor $\lambda^{c(s)}_{l=1} (k)$ and other values of $l$}
\label{sec_new}
\begin{table*}
  \begin{minipage}[t]{0.48\hsize}
    \begin{tabular}{lcccc}
    \hline\hline
    channel\quad& $\delta\chi_{l=1,L}$ & $\delta\chi_{l=1,M}$ & $\delta\chi_{l=1,H}$ & $\delta\chi_{l=1}$\\
    \hline
    charge & $-1$ & $+0.36 \pm 0.04$ &$+0.12 \pm 0.04$&  $-0.52\pm 0.06$ \\
    spin & $+1$ & $-0.86 \pm 0.02$ & $+1.50 \pm 0.04$&  $-0.36 \pm 0.04$ \\
    \hline\hline
  \end{tabular}
  \caption{Numerical results for high- and middle- energy contributions  to charge and spin susceptibilities for $l=1$ order parameters with the form factor  $\lambda\cs_{l=1} (p)=k_F {\bf p}/|{\bf p}|$ (``M'' and ``H'' terms), together with the analytical result for the low-energy ``L'' contribution. The numbers are in units of $\chib = m^3U^2 k^2_F/16\pi^3$. The results clearly deviate from those in Table \ref{tab:delta-chi} and do not satisfy Eqs.  \eqref{n22} and \eqref{n3}.
   \label{tab:delta-chi-nl}}
\end{minipage}
\hfill
\begin{minipage}[t]{0.48\hsize}
  \centering
  \begin{tabular}[c]{lcc}
    \hline\hline
    channel & from Eq.\eqref{eq:3vertex} & from $\delta\chi_{l=1,M}$\\
    \hline
    charge & $+1.07 \pm 0.01$ & $+1.07 \pm 0.01$\\
    spin & $-0.548 \pm 0.008$ & $-0.538 \pm 0.008$\\
    \hline\hline
  \end{tabular}
    \caption{Numerical results of $\Lambda_{l=1}\cs-1$ for the case when the form factor is $\lambda\cs_{l=1} (p)=k_F  {\bf p}/|{\bf p}|$.
     The numbers are in units of $\overline \Lambda = \frac{m^2U^2}{8\pi^2}$. The first column is obtained from a direct evaluation using Eq.\eqref{eq:3vertex}(in which ${\bf p}$ is replaced with $k_F  {\bf p}/|{\bf p}|$, and the second column is extracted from our calculation of $\delta\chi_{l=1,M}$ via Eq.\eqref{lambdaz}. The results show that  Eq.\eqref{n2} is not satisfied if the form-factor is different from ${\bf k}$.
  }
  \label{tab:delta-lambda-nl}
\end{minipage}
\end{table*}
The purpose of this final section is to clarify how generic are the constraints imposed by Eqs. \eqref{n2}, \eqref{n22}, which prevent a Pomeranchuk instability for charge-current and spin-current order parameters. In this section we first study the case of an order parameter $\rho\cs_{l=1}$ with form factor $\lambda_{l=1} (k) = \k f\cs_{l=1} (|k|)$ for which $f_{l=1}(|\k|) \neq 1$. We argue that in this case there is no relation $\Lambda\cs_{l}Z \propto (1 + F\cs_1)$ and therefore a Pomeranchuk instability does occur when $F_l\cs = -1$.

The argument is quite straightforward -- $\rho\cs_{l=1}$ with $f_{l=1}(|\k|) \neq 1$ is not a current of a conserved quantity, hence it is not related by a continuity equation to a quantity, such as a conserved charge, whose susceptibility is expressed in terms of Landau parameters. Rather, it has two pieces and is of the form,
\begin{equation}
  \label{eq:chi-fl-def}
  \chi_{l=1}\cs =  \tilde\chi_{l=1}\cs + \delta\chi\cs_{l=1},
\end{equation}
where $\tilde\chi_{l=1}\cs$ is finite and can be expressed in terms of Landau
parameters
at $\Omega/v_F |\q| \to 0$, but $\delta\chi_{l=1}\cs$ cannot. As a result,while $\tilde\chi_{l=1}\cs$ remains finite when $F_{l=1}\cs = -1$, $\delta\chi_{l=1}\cs$ diverges, signaling a Pomeranchuk instabilitiy.

An indication of this appears already at first order in $U$. To see this, we evaluate the diagrams of Fig. \ref{fig:chi-1st-order} in Sec. \ref{sec:first-order-uq} for the more general case $f_l(|\k|) \neq 1$. Then we find the contribution of diagrams \ref{fig:chi-1st-order}a,b is:
\begin{flalign}
 \label{eq:gen-chi-l1-1st-order}
 \left(\chi_{l=1,2a}\cs+\chi_{l=1,2b}\cs\right) = \frac{m^2\Omega^2}{q^2}\left(\tilde\chi_{l=0,2a}\cs+\tilde\chi_{l=0,2b}\cs\right)
 + \delta\chi_{l=1}\cs,
\end{flalign}
Here, $\tilde\chi_{l=0}\cs$ is the susceptibility of a channel with $l=0$ symmetry, but with $f_{l=1}(|\k|)$ in the form-factor, and
\begin{widetext}
\begin{align}
 \label{eq:chi-fl-1st}
 \delta\chi_{l=1}\cs &= 2\int\frac{d^2k d^2p}{(2\pi)^2}\frac{\left[n_F\left(\xi_{\p-\q/2}\right)-n_F\left(\xi_{\p+\q/2}\right)\right]\left[n_F\left(\xi_{\k-\q/2}\right)-n_F\left(\xi_{\k+\q/2}\right)\right]}{(m^{-1} q)^2} \nn \\
 &\qquad U(\p-\k)\left[f_{l=1}(|\k|)f_{l=1}(|\p|) - f_{l=1}^2(|\k|)\right]\left(1 - \frac{2\Omega}{\Omega - m^{-1}\k\cdot\q}\right),
\end{align}
\end{widetext}
is an additional term which is exactly zero for $f_l = 1$. The results to order $U$  are somewhat special because each of the three terms in Eq. \eqref{eq:gen-chi-l1-1st-order} has
an
additional $q^2$ factor in the $\Omega/q \to 0$ limit.
%

\begin{table*}
  \centering
  \begin{tabular}{lcccc}
    \hline\hline
    channel & $\delta\chi_{l=2,L}$ & $\delta\chi_{l=2,M}$ & $\delta\chi_{l=2,H}$ & $\delta\chi_{l=2}$\\
    \hline
    charge, $f_{l=2}=1$ & $+\frac{1}{2}$ & $-1.48 \pm 0.02$ & $+6.40 \pm 0.04$&  $\bf +4.92 \pm 0.04$ \\
    spin, $f_{l=2}=1$  & $-\frac{1}{2}$ & $-1.50 \pm 0.02$ & $+1.32 \pm 0.04$&  $\bf -0.18 \pm 0.04$ \\
    charge, $f_{l=2}=\frac{k_F^2}{|\k|^2}$ & $+\frac{1}{2}$ & $-1.46 \pm 0.02$ & $+1.18 \pm 0.04$&  $\bf -0.28 \pm 0.04$ \\
    spin, $f_{l=2}=\frac{k_F^2}{|\k|^2}$ & $-\frac{1}{2}$ &$-1.66 \pm 0.02$ & $+0.70 \pm 0.04$&  $\bf -0.96 \pm 0.04$ \\
    \hline\hline
  \end{tabular}
  \caption{Charge and spin susceptibilities in the quadrupolar $l=2$ channel, calculated from Eqs.\eqref{eq:all} using two different form factors. The numbers are in units of $\chib' = m^3U^2 k^4_F/16\pi^3$.  The results show no connection between high-energy and middle-energy contributions and the low-energy contribution.
  Different form factors depend on $f_{l=2}$ through our definition in Eq.\eqref{eq:form-factor-def}.}
  \label{tab:l2-data}
\end{table*}
Nevertheless, the  appearance of $ \delta\chi_{l=1}$ already at this order indicates that $\delta\chi\cs_{l=1}$ is not expressed via $\delta\chi\cs_{l=0}$, taken in the $q/\Omega \to 0$ limit, as it was the case for a current of a conserved order parameter.

To see explicitly that for $f_l \neq 1$ Eqs. \eqref{n2} and \eqref{n3} are no longer valid
we perform the same calculations as in Sec.  \ref{sec:micr-expl-absence}
for $f^{c,s}_{l=1} (|k|) \neq \text{constant}$. For definiteness, we consider $f\cs_{l=1}(|\k|) = k_F/|\k|$, i.e.,  $\lambda\cs_{l=1}(\k) = k_F\hat\k \cdot \hat \q$.
The cancellation between different diagrams for susceptibility still holds, and the results for $\delta \chi^s_{l=1}$ and $\delta \chi^s_{l=1}$ to order $U^2$ are still given by Eqs (\ref{ll3}) and (\ref{ll4}), and the contribution from each diagram can again be
split into ``H'', ``M'', and ``L'' parts. However,  now each contribution has to be computed with different prefactors. This does not affect the ``L'' contribution as, by construction, $f\cs_{l=1} (k_F) =1$, but the modification  of $f\cs_{l=1} (k)$ does affect ``M'' and ``H'' contributions.
In Table \ref{tab:delta-chi} we present the results for ``H'', ``M'', and ``L'' contributions to  $\delta \chi^c_{l=1}$ and  $\delta \chi^s_{l=1}$ in units of $\chib$. We also computed $\Lambda\cs_{l=1}$ by evaluating the renormalization of the three-leg vertex.  We show the results in Table \ref{tab:delta-lambda-nl}, again in units of $\overline{\Lambda}$.  We see that neither the constraints on the components of the susceptibilities, Eq.  (\ref{f1}), nor the conditions on $\Lambda\cs_{l=1}$,  Eq. \ref{eq:delta-gamma-z-expected}, are obeyed.
Therefore, $\Lambda\cs_{l=1}$ does not scale with $(1 + F\cs_1)$ and does not cancel $1/(1 + F\cs_1)$ in the quasiparticle part of the susceptibility. Since there is no cancellation of the diverging part,
a Pomeranchuk instability does occur when $F_1\cs = -1$ for any order parameter with $f\cs_{l=1} \neq 1$.

We also explicitly calculated ``L'', ``M'', and ``H'' contributions to susceptibility in $l=2$ with $f\cs_{l=2} (k) =1$ and  $f_{l=2}=\frac{k_F^2}{|\k|^2}$. For $l=2$, $F^c_2 = - F_2^s = \chib/2$, such that the low-energy contributions to the $l=2$ charge and spin susceptibilities are $\delta \chi^c_{l=2,L} = - \delta \chi^s_{l=2,L} = -\chib'/2$, where $\chib' = \chib k_F^2 = m^3U^2 k^4_F/16\pi^3$. We show the results in Table  \ref{tab:l2-data}.  We
didn't find
any relation between ``M'' and ``H'' contributions to both spin and charge susceptibilities and $1 + F\cs_{l=2}$. In particular, we checked the expressions for $l=2$ case presented in Ref. \onlinecite{Kiselev2017} and did not reproduce them. This can be also seen by comparing the results in Ref. \onlinecite{Kiselev2017} with our expressions for susceptibility to first order  in momentum-dependent $U(q)$, Eq. (\ref{n_12}).

\section{Summary}
\label{sec:6}

In this paper we studied the constraints placed by conservation laws on Pomeranchuk transitions, particularly the role of the continuity equation and longitudinal sum rule.  This issue has been
previously considered by Leggett \cite{Leggett1965} back in 1965, and was re-analyzed recently by Kiselev et al\cite{Kiselev2017}.
The continuity equation and the sum rule reveal interesting properties of susceptibilities of currents of conserved total charge and spin.
Namely, high energy features of a system, such as $\Lambda_{l=1}\cs Z$, and the incoherent piece of the susceptibility, $\chi\cs_{l=1,inc}$, can be expressed in terms of the Landau parameters $F_l\cs$, which describe the interaction between fermions on the FS. In particular, $\Lambda\cs_{l=1} Z$ scales as $(1+ F\cs_1)$ and vanishes at $F\cs_1 =-1$, when the quasiparticle contribution to susceptibility diverges as $1/(1+ F\cs_1)$. The vanishing of $\Lambda\cs_{l=1} Z$ cancels out the divergence, and, as a result, the system does not undergo a p-wave Pomeranchuk instability. Our aim was to verify this in diagrammatic perturbation theory, present a microscopic explanation  why high-energy and low-energy contributions to susceptibility are related, and check how general such constraints are.

We showed that the constraints work only for $l=1$ and for the specific
$l=1$ order parameter with form-factor $\lambda\cs_{l=1} (k) = {\bf k}$. Such an order parameter describes currents of the  fermionic number and spin  - both
of which
are conserved quantities. For any form factor with $l=1$ symmetry, but different functional behavior, $\lambda\cs_{l=1}(\k) = f\cs_{l=1}(|\k|) \k $ with $f(|\k|) \neq 1$, high-energy and low-energy contributions to the susceptibility are not correlated.  The same is true for other values of $l$.  As a result, the susceptibility for any other order parameter  with either $l=1$ or other $l$ diverges when $F\cs_l =-1$, i.e., the Pomeranchuk instability does occur.

\section{Acknowledgements}
  We thank  J. Schmalian, P. Woelfle, and particularly D. Maslov for  valuable discussions.  The work was supported by  NSF DMR-1523036.   
  AVC is thankful to KITP at UCSB  where part of the work was done. KITP is supported by NSF grant PHY-1125915.

\bibliography{ref}
\clearpage
\appendix
\section*{Appendix: Details of the numerical evaluation}
\label{sec:numerical_evaluation_details}
\begin{table*}[t]
  \caption{\label{tab:3vertex}Numerical results of high energy and mixed energy contributions for different form factors. The unit here are $k_F^2\frac{m^3U^2}{8\pi^3}$ for the first two lines and $k_F^4\frac{m^3U^2}{8\pi^3}$ for the last two lines.\label{tab:highandmixed}}
  \begin{ruledtabular}
  \begin{tabular}{ccccccc}
  Form factor&$\frac{1}{2}\chi_{l,6a}^H$ & $\chi_{l,6c}^H$ & $\frac{1}{2}\chi_{l,6d}^H$&$\frac{1}{2}\chi_{l,6a}^M$ & $\chi_{l,6c}^M$ & $\frac{1}{2}\chi_{l,6d}^M$\\
  \hline
  $\lambda_{p}=|\bm{p}|\cos\phi_{\bm{p}}$ & $0.613\pm0.008$ & $-0.226\pm0.017$ & $-0.379\pm0.005$ & $-0.879\pm0.008$ & $0.767\pm0.006$ & $ 0.608\pm0.005$\\

  $\lambda_{p}=k_F\cos\phi_{\bm{p}}$ & $0.500\pm0.013$ & $-0.193\pm0.008$ & $-0.250\pm0.008$ & $-0.879\pm0.008$ & $0.508\pm0.013$ & $ 0.548\pm0.005$\\

  $\lambda_{p}=|\bm{p}|^2\cos2\phi_{\bm{p}}$ & $1.956\pm0.015$ & $-0.052\pm0.008$ & $1.292\pm0.014$ & $-0.879\pm0.008$ & $0.266\pm0.006$ & $ -0.379\pm0.006$\\

  $\lambda_{p}=k_F^2\cos2\phi_{\bm{p}}$ & $0.500\pm0.013$ & $-0.058\pm0.006$ & $0.147\pm0.012$ & $-0.879\pm0.008$ & $0.202\pm0.008$ & $ -0.303\pm0.005$\\

  \end{tabular}
  \end{ruledtabular}
  \end{table*}
For our numerical evaluation of the high energy and middle energy contributions to the second order diagrams, Eqs. \eqref{eq:all} and \eqref{eq:3vertex} we used Mathematica
11.1.1 with the built-in algorithm \emph{NIntegrate}, using the Monte Carlo integration strategy. In our evaluation of diagrams we used polar coordinates and cut off the momentum at  $15k_F$, e.g. $\{|\p|,0,15k_F\}$.
The
UV divergence in Eqs. \eqref{eq:all} is avoided by
the
symmetry factor $\cos l\phi$ and this $15k_F$ truncation is large enough to obtain our results accurately. Since only the angle differences of three the momenta($\p$, $\k$ and $\k'$) enter our integrals, one can integrate out one of these three angles by hand to achieve higher accuracy.

In evaluations of the high energy contributions($\chi_{l,6a}^H$, $\chi_{l,6c}^H$ and $\chi_{l,6d}^H$), the integral region $\{|\p|,0,15k_F\}\times\{|\k|,0,15k_F\}\times\{|\k'|,0,15k_F\}$ is divided into 8 parts: every dimension of momentum is divided into $(0, 3k_F)$ and $(3k_F, 15k_F)$, e.g. $\{|\p|,0,15k_F\}=\{|\p|,0,3k_F\}+\{|\p|,3k_F,15k_F\}$ . In evaluations of mixed energy contributions($\chi_{l,6a}^M$, $\chi_{l,6c}^M$ and $\chi_{l,6d}^M$), the integral region $\{|\k|,0,15k_F\}\times\{|\k'|,0,15k_F\}$ is divided into 9 parts instead. Each momentum dimension is divided as $(0, 3k_F)$, $(3k_F, 6k_F)$ and $(3k_F, 15k_F)$.
Every subregion was sampled using a maximum
of $10^9$ points.
We evaluated each subregion 10 times to ensure the convergence of the numerical sums. The various $\chi_l\cs,\Lambda_l\cs Z$ we needed are readily found from the numerical expressions for the ``H'' and ``M'' diagrams as detailed in the text.  The deviation of these 10 evaluations are the basis for computing the error brackets of Tables \ref{tab:delta-chi}-\ref{tab:l2-data}.

As a check of the reliability of our numerical scheme we computed the quasiparticle residue $Z$, which is known to be $Z=1-1.39\frac{m^2U^2}{8\pi^2}$ (see text). Our calculation for $Z$ is based on Pitaevskii-Landau relations, Eq. \eqref{eq:Z-GI-FL} of the text and
\begin{equation}\label{eq:PL}
  \frac{1}{Z} = 1 -  \frac{i}{2}\sum_{\alpha\beta}\int\frac{d^3q}{(2\pi)^3}\Gamma^{\omega}_{\alpha\beta,\beta\alpha}(k_F\hat p,\q)(G^2_{q})^\omega
\end{equation}
Eq. \eqref{eq:PL} and Eq. \eqref{eq:Z-GI-FL} must give the same result. Numerically we found,
\begin{equation*}
	\begin{aligned}
		Z=1-(1.389\pm0.045)\frac{m^2U^2}{8\pi^2}\quad\text{based on Equation~\eqref{eq:PL}},\\
		Z=1-(1.390\pm0.028)\frac{m^2U^2}{8\pi^2}\quad\text{based on Equation~\eqref{eq:Z-GI-FL}},\\
	\end{aligned}
\end{equation*}
which gives us confidence our integrals are
accurate.

\end{document}